\documentclass[aps,groupedaddress, superscriptaddress,showpacs,amsmath,amssymb,twocolumn,pra]{revtex4-1}
\usepackage{dcolumn}
\usepackage{bm}
\usepackage[dvips]{graphicx}
\usepackage{wrapfig,subfigure}
\usepackage{amssymb}
\usepackage{color}
\usepackage[normalem]{ulem}
\usepackage{amsmath}
\usepackage{epstopdf}
\newcommand{\rb}{\boldsymbol{r}}
\newcommand{\nb}{\boldsymbol{n}}

\newcommand{\zb}{\boldsymbol{z}}
\newcommand{\Kb}{\boldsymbol{K}}

\begin{document}
\title{Correlators exceeding one in continuous measurements of superconducting qubits}
\author{Juan Atalaya}
\affiliation{Department of Electrical and Computer Engineering, University of California, Riverside, CA 92521, USA}
\author{Shay Hacohen-Gourgy}
\affiliation{Quantum Nanoelectronics Laboratory, Department of Physics, University of California, Berkeley CA 94720, USA}
\affiliation{Center for Quantum Coherent Science, University of California, Berkeley CA 94720, USA}
\affiliation{Department of Physics, Technion, Haifa 3200003, Israel}
\author{Irfan Siddiqi}
\affiliation{Quantum Nanoelectronics Laboratory, Department of Physics, University of California, Berkeley CA 94720, USA}
\affiliation{Center for Quantum Coherent Science, University of California, Berkeley CA 94720, USA}
\author{Alexander N. Korotkov}
\affiliation{Department of Electrical and Computer Engineering, University of California, Riverside, CA 92521, USA}
\date{\today}

\begin{abstract}
We consider the effect of phase backaction on the correlator $\langle I(t)\, I(t+\tau )\rangle$ for the output signal $I(t)$ from continuous measurement of a qubit. We demonstrate that the interplay between informational and phase backactions in the presence of Rabi oscillations can lead to the correlator becoming larger than 1, even though $|\langle I\rangle|\leq 1$. The correlators can be calculated using the generalized ``collapse recipe'' which we validate using the quantum Bayesian formalism. The recipe can be further generalized to the case of multi-time correlators and arbitrary number of detectors, measuring non-commuting qubit observables. The theory agrees well with experimental results for continuous measurement of a transmon qubit. The experimental correlator exceeds the bound of 1 for a sufficiently large angle between the amplified and informational quadratures, causing the phase backaction. The demonstrated effect can be used to calibrate the quadrature misalignment.
\end{abstract}

\maketitle
{\it Introduction.} Continuous quantum measurements (CQMs) are attracting significant attention in quantum computing and quantum physics. Although they have been theoretically discussed for a long time using various approaches~\cite{BookKraus, Diosi1988, Dalibard1992, Belavkin1992, BookCarmichael, WisemanMilburn1993, Korotkov1999, Gambetta2008, Korotkov2011},  current interest in CQMs is mainly motivated by relatively recent experiments with superconducting qubits~\cite{Katz2006partial-collapse, Laloy2010, Murch2013, Hatridge2013, Riste2013, Shay2016, Huard2017}. They are useful for quantum computing applications such as quantum feedback~\cite{Wiseman1993,Ruskov2002,Vijay2012,Lange2014,Patti2017}, rapid state purification~\cite{Jacobs2003}, preparation of entangled states~\cite{Ruskov2003,Riste2013,Roch2014}, and continuous quantum error correction~\cite{Landahl2002, Mabuchi2009}. CQMs are also shedding light on our understanding of the still debatable quantum measurement process, including nontrivial cases such as simultaneous CQM of noncommuting observables~\cite{Ruskov2010, Shay2016, Huard2017}.

Temporal correlators of the output signals from CQMs are important objects to study because they bear nonclassical features due to the interplay between coherent quantum evolution and measurement-induced quantum backaction. In particular, violation of a classical bound is a clear indication of quantum behavior. As an example, macrorealism assumptions have been tested with correlators from CQM via the continuous Leggett-Garg inequality~\cite{Laloy2010}. There is significant recent interest in correlators from CQMs \cite{Murch2016, Diosi2016, Atalaya2018npj, Areeya2017, Atalaya2018multi, Tilloy2018}, including multi-time correlators and the case of non-commuting observables. In particular, multi-time correlators are important in the continuous operation of quantum subsystem codes \cite{Atalaya2017QEC}.

Quantum backaction from measurement can be described in terms of Kraus operators~\cite{BookKraus}. The polar decomposition of a Kraus operator suggests, in general, two types of quantum backaction that are related to the non-unitary and unitary factors of the polar decomposition. In particular, in circuit QED-based measurements of superconducting qubits they are often referred to as informational backaction and phase backaction, respectively~\cite{Korotkov2011,Murch2013,Korotkov2016}. Circuit QED systems are ideal to study these two types of quantum backaction because their relative strength is easily tunable by the phase of the pump applied to a phase-sensitive parametric amplifier~\cite{Gambetta2008, Korotkov2011, Murch2013}.

In this paper, we study the effect of phase backaction on output-signal correlators for continuous measurement of a superconducting qubit.
We present a general theory for multi-time correlators in the spirit of the ``collapse recipe''~\cite{Korotkov2001sp, Atalaya2018npj, Atalaya2018multi}, which is extended here to include phase backaction and proven using the quantum Bayesian formalism. In such a generalized recipe, the correlators from continuous qubit measurements can be calculated by assuming fictitious ``strong'' measurements (with discrete outcomes $\pm1$) at the time moments entering the correlator and assuming ensemble-averaged evolution at other times. Importantly, the fictitious strong measurements can move the qubit state {\it outside the Bloch sphere}, and correspondingly the outcome probabilities for the next strong measurement can be negative. Even though the procedure is bizarre from physical point of view, this is a simple way to obtain correct correlators, including the case of simultaneous CQM of noncommuting qubit observables and arbitrary additional evolution and decoherence of the qubit.

In particular, our theory predicts the counterintuitive result that {\it correlators can be larger than} 1, even though the average value of the output is between $\pm 1$. To test this prediction, we perform CQM of $\sigma_z$ (Fig.\ 1) and show that the experimental correlators indeed exceed unity when we use a sufficiently strong phase backaction and sufficiently fast Rabi oscillations.
Note that such non-classical values would be natural for weak values \cite{AAV1988}; however, our experiment is not related to weak values since it does not use post-selection. We also discuss a sensitive correlator-based method to estimate the misalignment between amplified and informational quadratures in circuit QED-based qubit measurement setups.

\begin{figure}[t!]
\centering
\includegraphics[width=0.9\linewidth, trim = 2.5cm 0.5cm 2cm 1.6cm, clip=true]{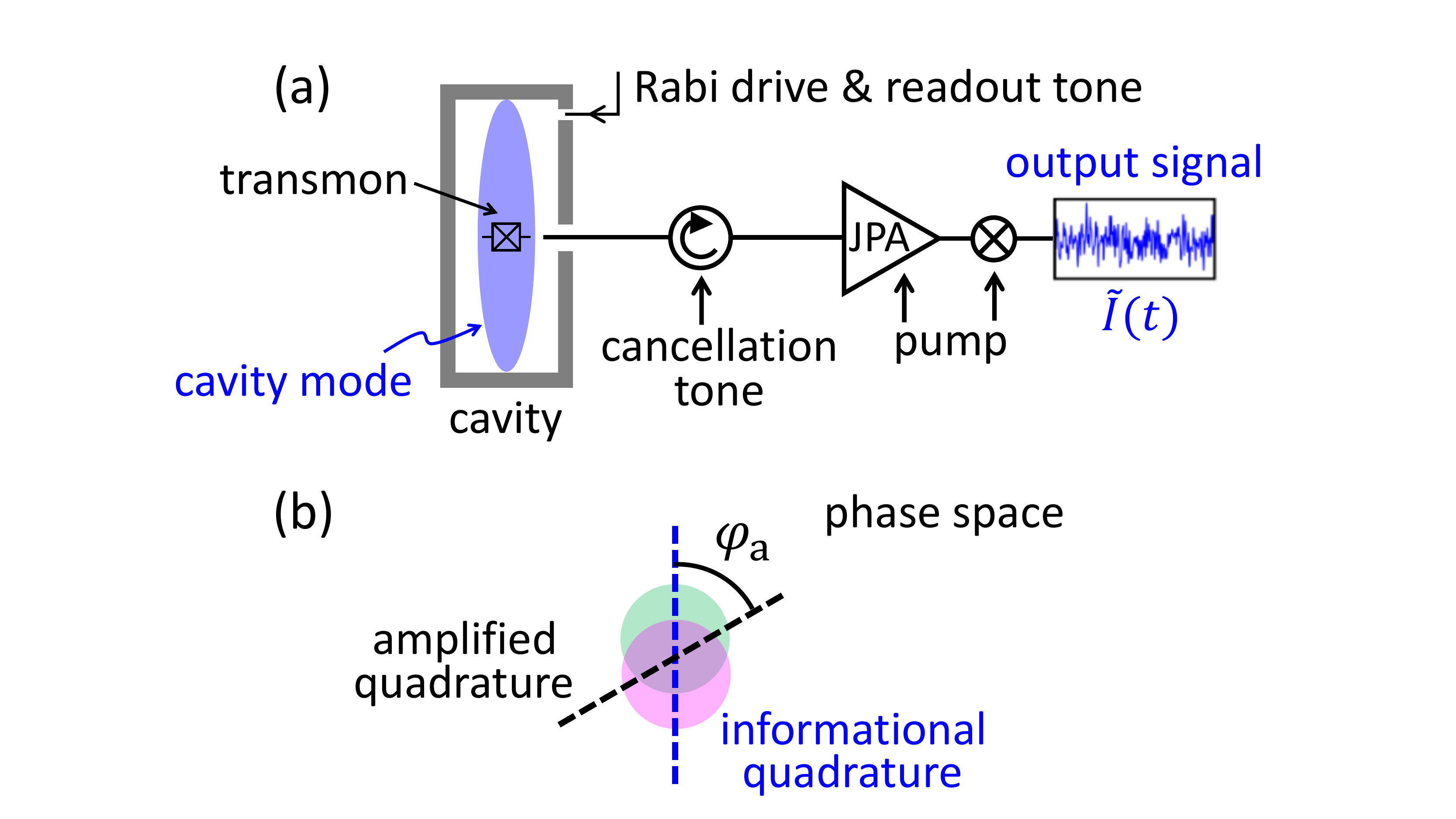}
\caption{(a) Schematic illustration of the experimental setup for continuous measurement of qubit observable $\sigma_z$. A superconducting qubit is dispersively coupled to the fundamental mode of a 3D microwave resonator. The  leaked field is amplified by a phase-sensitive Josephson parametric amplifier (JPA), producing the (downconverted) normalized output signal $I(t)$. The cancellation tone displaces the outgoing field close to the vacuum, thus preventing JPA saturation.
The coherent states corresponding to the eigenstates of $\sigma_z$ are illustrated in panel (b) by two circles in phase space. The line through their centers defines the informational quadrature, while the JPA's pump phase defines the amplified quadrature. The angle $\varphi_{\rm a}$ between them affects the phase backaction.
 }
\label{fig:setup}
\end{figure}
%

{\it The quantum Bayesian formalism.} As the simplest case, let us consider a Rabi-rotated qubit under continuous $\sigma_z$-measurement in the typical circuit QED setup with a phase-sensitive amplifier \cite{Murch2013,Shay2016} -- see Fig.\ 1. In this case the relative strength of the phase backaction and informational backaction is controlled by the angle $\varphi_{\rm a}$ between the amplified and informational quadratures \cite{Gambetta2008, Korotkov2011}. We will discuss the correlator ($t_2>t_1$)
\begin{equation}
\label{eq:Kzz-def}
K(t_1,t_2) \equiv \langle I(t_2)\, I(t_1)\rangle,
\end{equation}
where $I(t)=[\tilde I(t)-\tilde I_{\rm o}]/\Delta I(\varphi_{\rm a})$ is the normalized output signal, $\tilde{I}(t)$ is the actual experimental output, $\tilde I_{\rm o}$ is the offset, and $\Delta I(\varphi_{\rm a})=\Delta I_{\rm max} \cos \varphi_{\rm a}$ is the response,  so that this normalization provides $\langle I\rangle=1$ or $-1$ when the qubit is in the state $|1\rangle$ or $|0\rangle$, respectively (the symbol $\langle ..\rangle$ means ensemble average). The normalized signal can be modeled as ~\cite{Korotkov1999,  Atalaya2018multi}
\begin{equation}
\label{eq:Iz}
I(t)  = {\rm Tr}[\sigma_z\rho(t)] + \sqrt{\tau_{\rm m}}\,\xi(t)  = \zb\boldsymbol{r}(t) + \sqrt{\tau_{\rm m}}\, \xi(t),
\end{equation}
where $\boldsymbol{r}=(x,y,z)$ is the Bloch vector defined by the qubit density matrix parametrization $\rho = (\openone + x\sigma_x + y\sigma_y + z\sigma_z)/2$ and $\zb=(0,0,1)$ is the measurement axis direction corresponding to the measured observable $\sigma_z=|1\rangle \langle 1|-|0\rangle \langle 0|$. The white Gaussian noise $\xi (t)$ has zero average, $\langle \xi(t)\rangle=0$, and two-time correlator
\begin{equation}
\label{eq:xi-corr}
\langle\xi(t)\,\xi(t')\rangle=\delta(t-t').
\end{equation}
The ``measurement time'' $\tau_{\rm m}=\tau_{\rm min}/\cos^{2} \varphi_{\rm a}$ in Eq.~\eqref{eq:Iz} is the time to reach  the signal-to-noise ratio of 1.

The qubit evolution can be described by the quantum Bayesian equation \cite{Gambetta2008,Korotkov2011} (in It\^o interpretation)
\begin{equation}
\label{eq:QBE}
\dot\rb = \Lambda_{\rm ens}(\rb -\rb_{\rm st})+ \frac{\zb - (\zb \rb)\, \rb}{\sqrt{\tau_{\rm m}}}\,\xi(t) +\mathcal{K}\,\frac{ \zb \times \rb}{\sqrt{\tau_{\rm m}}}\,\xi(t),
\end{equation}
where the first term is the ensemble-averaged evolution, the second term is the informational backaction, and the third term is the phase backaction with $\mathcal{K}=\tan\varphi_{\rm a}$. The evolution of the ensemble-averaged state $\rb_{\rm ens}\equiv\langle \rb\rangle$,
    \begin{equation}
\label{eq:ens-avg-evol}
\dot \rb_{\rm ens} = \Lambda_{\rm ens}(\rb_{\rm ens} - \rb_{\rm st}),
\end{equation}
is characterized by $3\!\times\! 3$ matrix $\Lambda_{\rm ens}$ and stationary state $\boldsymbol{r}_{\rm st}$; this evolution corresponds to the  Lindblad-form equation, $\dot \rho_{\rm ens} = -(i/\hbar)[H_{\rm q},\rho_{\rm ens}] + \mathcal{L}[\rho_{\rm ens}]$, where $H_{\rm q}$ is the qubit Hamiltonian and $\mathcal{L}$ describes the qubit ensemble decoherence. In our case, the contribution to $\mathcal{L}$ due to measurement is $\mathcal{L}_{\rm m}[\rho] = \Gamma_{\rm m}[\sigma_z \rho\sigma_z - \rho]/2$, where $\Gamma_{\rm m}=(1+\mathcal{K}^2)/(2\eta\tau_{\rm m})=1/(2\eta\tau_{\min})$ is the measurement-induced ensemble dephasing rate and $\eta$ is the detector quantum efficiency. Note that $\Gamma_{\rm m}$ does not depend on $\varphi_{\rm a}$, in contrast to $\mathcal{K}$ and $\tau_{\rm m}$.

{\it Collapse recipe.} The collapse recipe was previously introduced to calculate two-time correlators  \cite{Korotkov2001sp} and multi-time correlators~\cite{Atalaya2018multi} without phase backaction. For the correlator (\ref{eq:Kzz-def}), this recipe states that we should replace continuous measurement at time moments $t_1$ and $t_2$ by (fictitious) projective measurements and use ensemble-averaged evolution at any other time. The projective measurements probabilistically produce discrete results $I_k=\pm 1$ and correspondingly collapse the qubit to $|1\rangle$ or $|0\rangle$.

As will be proven below, in the presence of phase backaction, the correlator (\ref{eq:Kzz-def}) still can be calculated in a somewhat similar way; however, we should use a quite unusual Generalized Collapse Recipe (GCR).
In particular, after a projective measurement at time $t_1$ with the result $I_1=\pm 1$, the qubit state collapses to $I_1 \rb_{\rm coll}$, where
\begin{equation}
\label{eq:rcoll}
\rb_{\rm coll} = \zb + \mathcal{K}\, ( \zb \times \rb_1)
\end{equation}
and $\rb_1\equiv \rb(t_1-0)$ is the qubit state just before the collapse.  We emphasise that, excluding the case when  $\zb \times \rb_1={\mathbf0}$ or $\mathcal{K}=0$, state~\eqref{eq:rcoll} is {\it outside the Bloch sphere}. After the collapse at time $t_1$, the qubit evolves according to  Eq.~\eqref{eq:ens-avg-evol}. Thus, using the GCR, the  correlator~\eqref{eq:Kzz-def} can be calculated as
\begin{equation}
\label{eq:coll-recipe}
K(t_1,t_2) = \sum_{I_{1},I_2=\pm1}I_1\,I_2\,p\big(I_2,t_2\big|I_1,t_1\big)\,p\big(I_1,t_1\big),
\end{equation}
where the sum is over four scenarios of outcomes,
\begin{equation}
\label{eq:pI1}
p\big(I_1,t_1\big) = \frac{1+I_1\,\zb\rb_1}{2}
\end{equation}
is the probability to get the first outcome $I_1=\pm 1$, and
\begin{equation}
\label{eq:pI2-I1}
p\big(I_2,t_2\big|I_1,t_1\big) = \frac{1+I_2\,\zb\rb_{\rm ens}\big(t_2\big|I_1\rb_{\rm coll},t_1\big)}{2},
\end{equation}
is the ``conditional probability'' to get the outcome $I_2=\pm 1$ at time $t_2$ given that we got outcome $I_1$ at time $t_1$. Here  $\rb_{\rm ens}\big(t\big|\rb_{\rm in},t_{\rm in}\big)$ denotes the solution of Eq.~\eqref{eq:ens-avg-evol} with initial condition $\rb_{\rm ens}(t_{\rm in})= \rb_{\rm in}$ at time $t_{\rm in}< t$. Since $\rb_{\rm ens}$ can be outside the Bloch sphere, the  ``probability'' \eqref{eq:pI2-I1} can be negative or larger than one; however, the normalization condition
$\sum_{I_2=\pm1} p\big(I_2,t_2\big|I_1,t_1\big)=1$
%
still holds. If the qubit is prepared in the state $\rb_{\rm 0}$ at $t_{\rm 0}<t_1$, then $\rb_1= \rb_{\rm ens}\big(t_1\big|\rb_{\rm 0},t_{\rm 0}\big)$ is within the Bloch sphere, so the first probability~\eqref{eq:pI1} has the usual range of values. Note that the recipe for multi-time correlators (discussed below) has essentially the same form.

{\it GCR from the quantum Bayesian formalism.} Let us prove the recipe of Eqs.~\eqref{eq:rcoll}--\eqref{eq:pI2-I1} using Eqs.~\eqref{eq:Iz}--\eqref{eq:ens-avg-evol}. The proof somewhat follows Refs.~\cite{Atalaya2018npj,Atalaya2018multi}. First, we rewrite Eq.~\eqref{eq:coll-recipe} of the GCR as
\begin{eqnarray}
\label{eq:Kzz-expanded}
&& K(t_1, t_2)=\zb\, \big[  \rb_{\rm ens}\big(t_2\big|\rb_{\rm coll},t_1\big) \, ( 1 + z_1 )/2
 \nonumber\\
&& \hspace{2.1cm} -  \rb_{\rm ens}\big(t_2\big|-\rb_{\rm coll},t_1\big) \, (1 - z_1)/2 \big],
\end{eqnarray}
where $z_1\equiv \zb\rb_1$ and $t_2>t_1$. Next, we calculate the correlator (\ref{eq:Kzz-def}) directly and show that the result coincides with Eq.\ (\ref{eq:Kzz-expanded}). Using Eq.\ (\ref{eq:Iz}), we decompose the  correlator as
\begin{equation}
\label{eq:Kzz-decom}
K(t_1,t_2) = \zb \, [ {\boldsymbol K}^{(1)}(t_1,t_2) + \boldsymbol{K}^{(2)}(t_1,t_2)],
\end{equation}
where the vector-valued correlators $\Kb^{(1,2)}$ are defined as
\begin{equation}
\label{eq:K1-K2}
{\boldsymbol K}^{(1)}(t_1,t_2) \equiv \langle \rb(t_2)\rangle \, z_1,\,{\boldsymbol K}^{(2)}(t_1,t_2)\equiv\langle \rb(t_2)\sqrt{\tau_{\rm m}}\,\xi(t_1)\rangle.
\end{equation}

Differentiating $\Kb^{(1)}$ over $t_2$ and using Eq.~\eqref{eq:QBE},  we find  that $\Kb^{(1)}$ satisfies an equation similar to Eq.~\eqref{eq:ens-avg-evol},
\begin{equation}
\label{eq:K1-eqm}
\partial_{t_2}{\boldsymbol K}^{(1)} (t_1, t_2) = \Lambda_{\rm ens}[{\boldsymbol K}^{(1)}(t_1, t_2) - z_1\rb_{\rm st}],
\end{equation}
with initial condition $\Kb^{(1)}(t_1,t_1) = \rb_1z_1$. Therefore,
\begin{equation}
\label{eq:K1-sol}
{\boldsymbol K}^{(1)}(t_1,t_2) = \mathcal{P}(t_2|t_1)\, z_1\rb_1 + z_1\boldsymbol{\mathcal{P}}_{\rm st}(t_2|t_1),
\end{equation}
where $\mathcal{P}(t|t')$ is a $3\!\times\! 3$ matrix satisfying  equation $\partial_t \mathcal{P}(t|t') = \Lambda_{\rm ens}(t)\,\mathcal{P}(t|t')$ with $\mathcal{P}(t'|t')=\openone$, and  $\boldsymbol{\mathcal{P}}_{\rm st}(t|t') = -\int_{t'}^t \mathcal{P}(t|t'')\, \Lambda_{\rm ens}(t'')\, \rb_{\rm st}(t'')\, dt''$ is a vector.

Similarly, ${\boldsymbol K}^{(2)}$ satisfies equation
\begin{equation}
\label{eq:K2-eqm}
\partial_{t_2} {\boldsymbol K}^{(2)} (t_1, t_2) = \Lambda_{\rm ens} {\boldsymbol K}^{(2)} (t_1, t_2),
\end{equation}
in which there is no term proportional to $\rb_{\rm st}$, in contrast to Eq.~\eqref{eq:K1-eqm}, because $\langle\Lambda_{\rm ens}\,\rb_{\rm st}\,\xi(t)\rangle=0$. To find the initial condition ${\boldsymbol K}^{(2)}(t_1,t_1+0)$, we discretize Eq.~\eqref{eq:QBE} with a timestep $\delta t$ and obtain $\rb(t_1+\delta t) -\rb(t_1)\approx [\zb  - z_1\rb_1 + \mathcal{K}\, (\zb \times\rb_1)]\,\delta t\,\xi(t_1)/{\sqrt\tau_{\rm m}}$, which has a typical size $\sim \sqrt{\delta t}$ since $\langle \xi^2(t_1)\rangle =(\delta t)^{-1}$ -- see Eq.~\eqref{eq:xi-corr}. Inserting this result for $\rb(t_1+\delta t)$ into Eq.~\eqref{eq:K1-K2}, we obtain $\Kb^{(2)}(t_1,t_1+0) = \zb  - z_1\rb_1 + \mathcal{K}\, (\zb \times \rb_1)$ in the limit $\delta t\to 0$. Thus,
\begin{equation}
\label{eq:K2-sol}
{\boldsymbol K}^{(2)}(t_1,t_2) = \mathcal{P}(t_2|t_1)\big[\rb_{\rm coll} - z_1\rb_1 \big],
\end{equation}
so that $\tau_{\rm m}$ in the definition \eqref{eq:K1-K2} of $\Kb^{(2)}$ cancels out.

From Eqs.~\eqref{eq:Kzz-decom},~\eqref{eq:K1-sol} and~\eqref{eq:K2-sol}, we obtain
\begin{equation}
\label{eq:Kzz-expanded-v2}
K(t_1,t_2) = \zb \big[\mathcal{P}(t_2|t_1)\, \rb_{\rm coll}+ z_1\boldsymbol{\mathcal{P}}_{\rm st}(t_2|t_1)\big],
\end{equation}
with the terms proportional to $z_1\rb_1$ in Eqs.~\eqref{eq:K1-sol} and~\eqref{eq:K2-sol} exactly cancelling each other and not contributing to Eq.~\eqref{eq:Kzz-expanded-v2}. This is expected from linearity of quantum mechanics, which requires a linear (not quadratic) dependence of the correlators on the initial state.

Finally, formally solving  Eq.~\eqref{eq:ens-avg-evol} as $\rb_{\rm ens}\big(t\big| \rb_{\rm in},t_{\rm in}\big) = \mathcal{P}(t|t_{\rm in})\, \rb_{\rm in} + \boldsymbol{\mathcal{P}}_{\rm st}(t|t_{\rm in})$ and using this solution in Eq.~\eqref{eq:Kzz-expanded}, we see that the result exactly coincides with Eq.~\eqref{eq:Kzz-expanded-v2}. This proves that the GCR yields the same correlator as the one obtained from the quantum Bayesian formalism.

{\it Experimental correlators larger than 1.} The GCR introduces an unusual way of thinking about the qubit evolution that nevertheless enables us to calculate correlators in CQMs. Next we discuss that the effective qubit evolution outside the Bloch sphere leads to correlators larger than 1 in the experiment illustrated in Fig.\ 1. In the experiment
the qubit undergoes Rabi oscillations with frequency $\Omega_{\rm R}$ over the $x$-axis and continuous measurement of $\sigma_z$. Neglecting the energy relaxation, the ensemble-averaged evolution is described by Eq.\ (\ref{eq:ens-avg-evol}) with $\rb_{\rm st}=0$ (i.e., unital evolution) and
\begin{equation}
\label{eq:Lambda-ens}
\Lambda_{\rm ens} =\left( \begin{array}{ccc} -\Gamma & 0 & 0 \\ 0 &-\Gamma & -\Omega_{\rm R} \\ 0& \Omega_{\rm R}& 0 \end{array}\right),
\end{equation}
where $\Gamma$ is the ensemble dephasing rate, which is mostly due to measurement, $\Gamma\approx \Gamma_{\rm m}$.
Because of unitality ($\rb_{\rm st}=0$), there is a symmetry
\begin{equation}
\label{eq:unital-cond}
\rb_{\rm ens}\big(t\big| - \rb_{\rm in},t_{\rm in}\big) = - \rb_{\rm ens}\big(t\big|\, \rb_{\rm in},t_{\rm in}\big),
\end{equation}
so Eq.\ \eqref{eq:Kzz-expanded} for the correlator   reduces to
    \begin{equation}
K(t_1,t_2)=\zb \,\rb_{\rm ens}\big(t_2\big|\rb_{\rm coll},t_1\big).
    \end{equation}
Therefore, in the GCR we can assume that the measurement result at $t_1$ is always $I_1=+1$, this moves the qubit to the state $\rb_{\rm coll}$ given by Eq.\ (\ref{eq:rcoll}), and the correlator is simply the qubit $z$-component at time $t_2$, i.e., $K=z_{\rm ens}\equiv \zb \rb_{\rm ens}$.

\begin{figure}[t!]
\centering
\begin{tabular}{cc}
\includegraphics[width=0.65\linewidth, trim =8cm 3cm 8cm 3cm, clip=true]{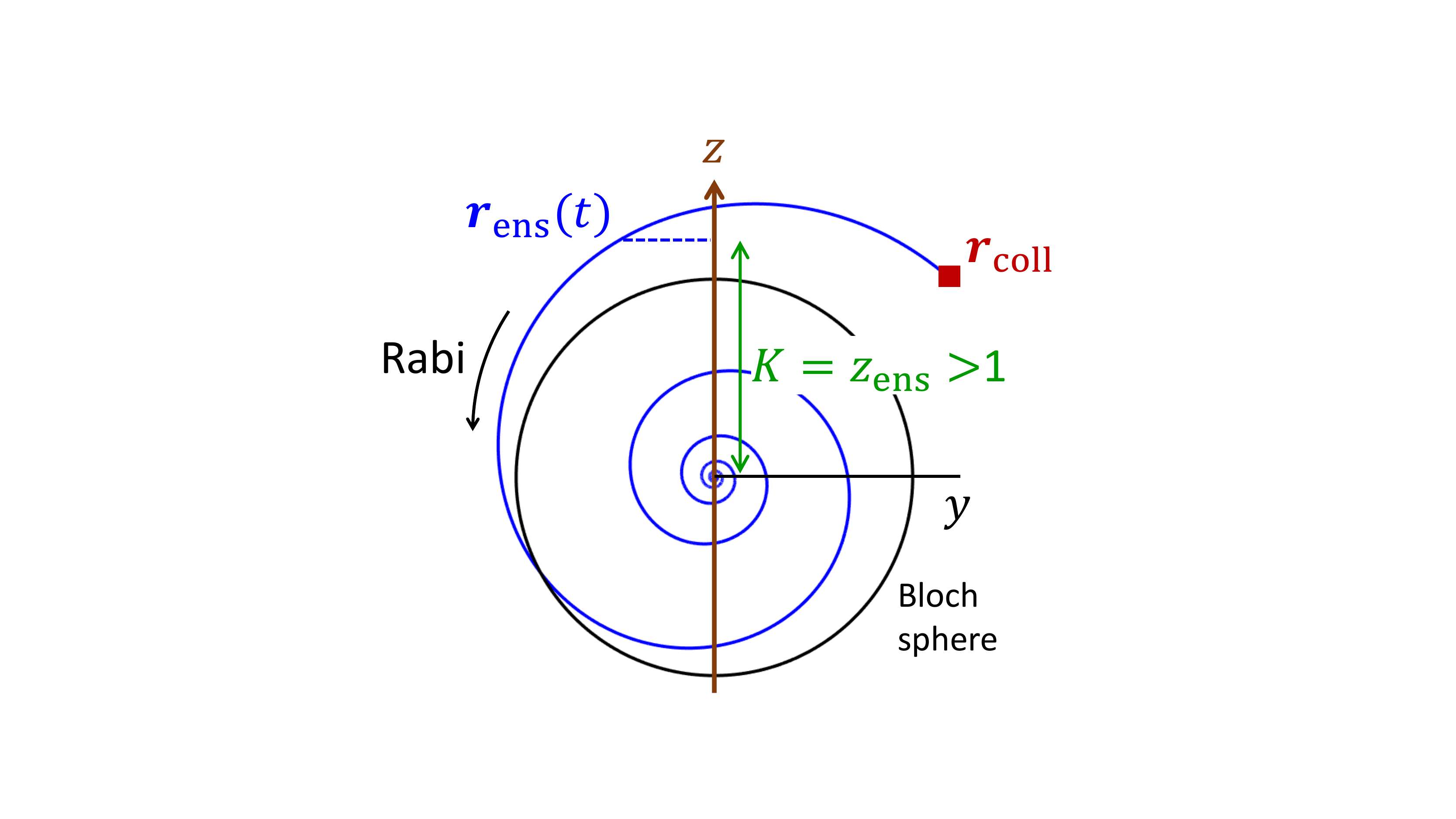}
\end{tabular}
\caption{Qubit evolution in the GCR picture. At time $t_1$, the qubit state jumps to $\rb_{\rm coll}$ [Eq.~\eqref{eq:rcoll-example}], which is outside the Bloch sphere when phase backaction is present. Rabi oscillations then can produce $z$-component $z_{\rm ens}\equiv \zb\rb_{\rm ens}$ larger than 1, so that the correlator $K(t_1,t_2)=z_{\rm ens}(t_2)$ exceeds 1.
\vspace{-0.1cm} }
\label{fig:fig2}
\end{figure}

In the experiment, the qubit is prepared at time $t_{\rm 0}=0$ in the state  $\rb_{\rm 0}=(x_{\rm 0},0,0)$ with $x_{\rm 0}=\pm 1$ (i.e., along the rotation axis). Without the intuition provided by the GCR, this choice to observe correlators larger than 1 is counterintuitive. However, according to the GCR, the effective after-collapse qubit evolution starts outside the Bloch sphere at the state
\begin{equation}
\label{eq:rcoll-example}
\rb_{\rm coll} = (0,x_1\tan\varphi_{\rm a},1),\;\;\;x_1 = x_{\rm 0}\exp({-\Gamma t_1}),
\end{equation}
which after Rabi rotation can have $z$-component up to $\sqrt{1+x_1^2 \tan^2 \varphi_{\rm a}}$. This geometrical picture is illustrated in Fig.~\ref{fig:fig2}, making  clear that both phase backaction and Rabi oscillations are necessary to observe $K>1$.

In the experiment, the correlator is additionally time-averaged in order to reduce fluctuations,
\begin{equation}
\label{eq:expt-corr}
K(\tau) \equiv \frac{1}{T} \int_{t_{\rm skip}}^{t_{\rm skip}+T}  K(t_1,t_1+\tau)\,{dt_1},
\end{equation}
where $T$ is the averaging duration, which starts with a small delay $t_{\rm skip}$ to skip initial transients.
Using the GCR, we obtain -- see the Supplemental Material (SM) ~\cite{SM},
\begin{eqnarray}
\label{eq:Kzz-result}
K(\tau) &=&\,  \left [\cos (\tilde\Omega_{\rm R}\tau) + \frac{\Gamma}{2\tilde\Omega_{\rm R}}\,\sin(\tilde\Omega_{\rm R} \tau)\right]e^{-\Gamma\tau/2}
    \nonumber \\
&&  +\tan\varphi_{\rm a}\,\frac{c\,x_{\rm 0}\,\Omega_{\rm R}}{\tilde\Omega_{\rm R}}\, \sin(\tilde\Omega_{\rm R} \tau)\, e^{-\Gamma\tau/2},
\end{eqnarray}
where $c = \exp(-\Gamma t_{\rm skip}) [1-\exp(-\Gamma T)]/(\Gamma T)$ and $\tilde\Omega_{\rm R}\equiv \sqrt{\Omega_{\rm R}^2 - \Gamma^2/4}$. This correlator does not depend on the quantum efficiency $\eta$. The first and second terms in Eq.~\eqref{eq:Kzz-result} are due to informational and phase backactions, respectively. Note that the quantum regression formula~\cite{GardinerBook} applied to the qubit state
gives only the first term ~\cite{Atalaya2018npj} and cannot be used in the case with phase backaction. Though theoretically $K(\tau)$ can exceed unity for any non-zero values of $\Omega_{\rm R} $ and $\varphi_{\rm a}$, in the experiment we need sufficiently fast Rabi oscillations and rather large $\varphi_{\rm a}$ to overcome experimental fluctuations. From Eq.~\eqref{eq:Kzz-result} for $|\Omega_{\rm R}|\gg\Gamma$, the maximum value of $K(\tau)$ is $K_{\max}\approx \sqrt{1 + c^2\tan^2\varphi_{\rm a}}$.

\begin{figure}[t!]
\centering
\begin{tabular}{c}
\includegraphics[width=0.85\linewidth, trim =1.5cm 4.5cm 0cm 0cm, clip=true]{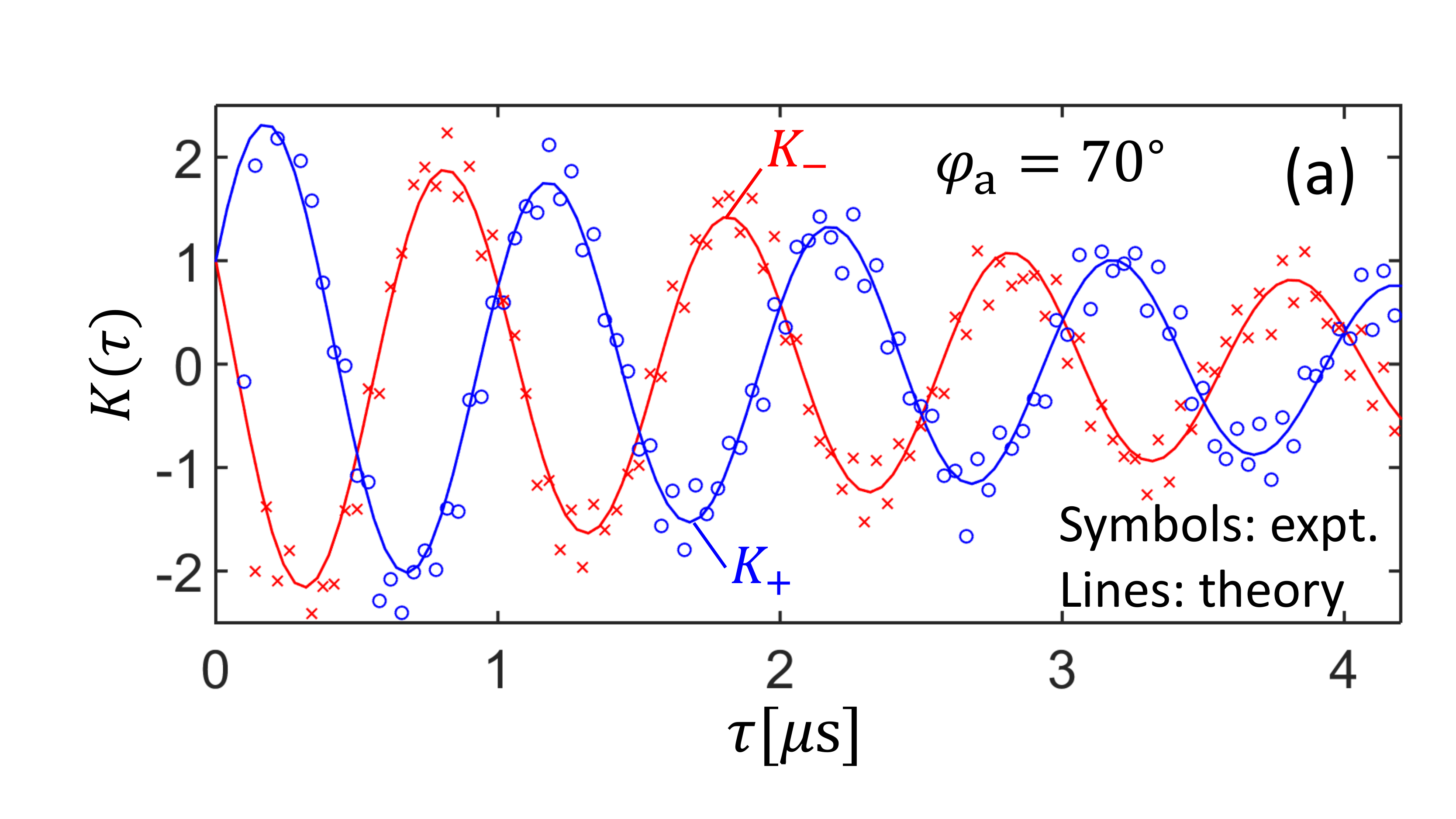} \\
\includegraphics[width=0.85\linewidth, trim =1.5cm 1.2cm 0cm 1.cm, clip=true]{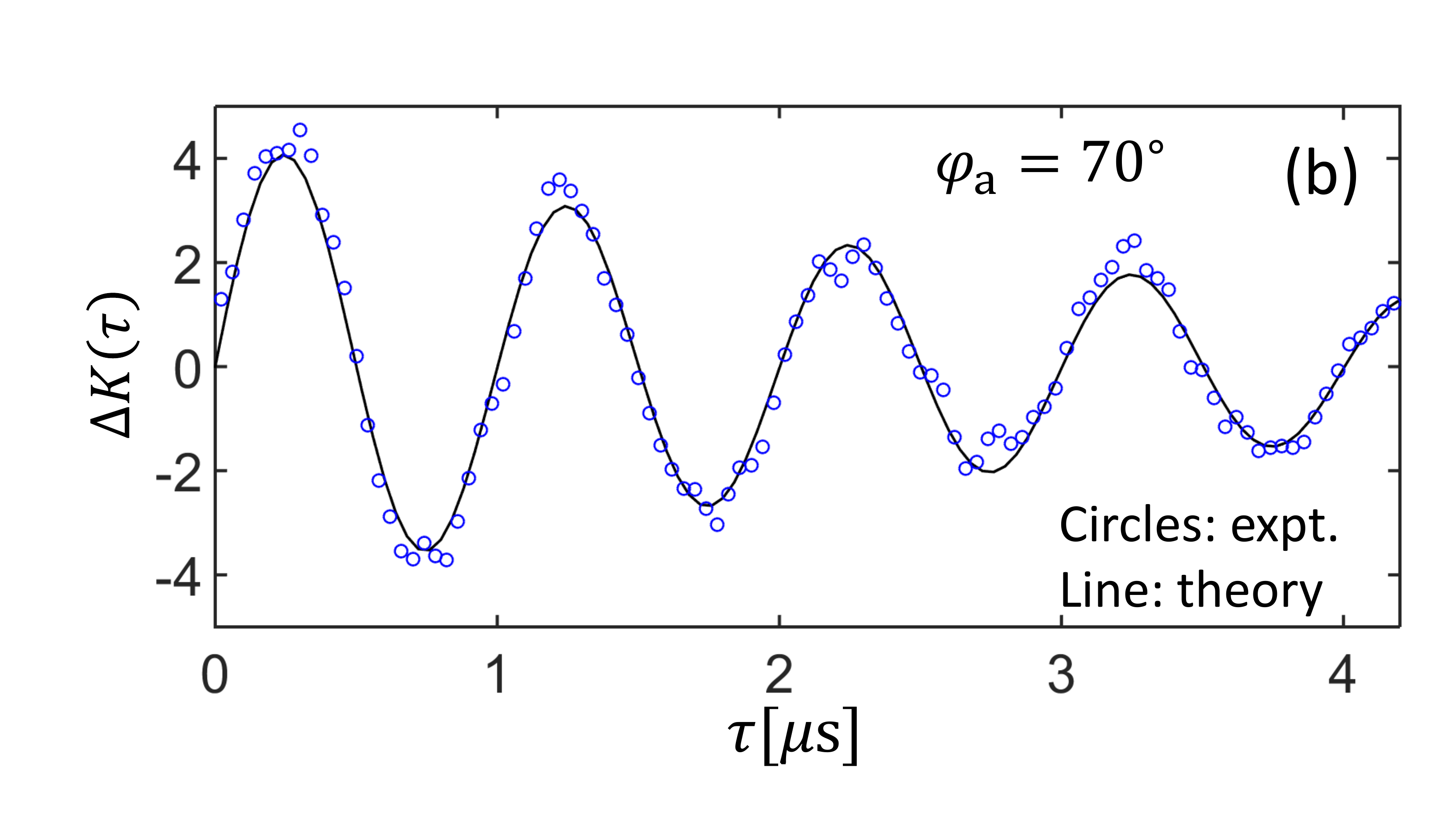}
\end{tabular}
\caption{Experimental correlators exceeding unity, for the phase misalignment $\varphi_{\rm a}=70^\circ$,  initial state $x_0=\pm 1$, Rabi frequency $\Omega_{\rm R}/2\pi=\pm 1$~MHz, and ensemble dephasing $\Gamma =1/1.8\,\mu$s. Panel (a) shows the correlators $K_\pm$, with $\pm$ corresponding to the sign of $x_{\rm 0}\Omega_{\rm R}$. Panel (b) shows the correlator difference $\Delta K(\tau) = K_+(\tau) - K_-(\tau)$. Experimental results are represented by symbols, the theory is shown by lines.
\vspace{-0.1cm} }
\label{fig:fig3}
\end{figure}
%


The measurement setup is shown in Fig.~\ref{fig:setup} and further discussed in the SM~\cite{SM}. In the experiment we use $\Gamma =1/1.8\,\mu$s, $\Omega_{\rm R}/2\pi=\pm 1$~MHz, and $\varphi_{\rm a}=70^\circ$. (In the SM \cite{SM}, we also present data for $\varphi_{\rm a}=0, \, 40^\circ$, and $80^\circ$.) The averaged correlator~\eqref{eq:expt-corr} is obtained from the recorded data using $T= 0.28\,\mu$s and  $t_{\rm skip}= 0.28\,\mu$s, so that $c= 0.79$ in Eq.~\eqref{eq:Kzz-result}. Figure~\ref{fig:fig3}(a) shows the experimental correlators $K_\pm(\tau)$, where the subscript indicates the sign of the product $x_{\rm 0}\Omega_{\rm R}$ \cite{SM}. In each case the ensemble averaging is over $6.5\times10^5$ recorded traces. We see a good agreement between experiment (symbols) and theory (lines) in Fig.\ 3(a). Most importantly, experimental correlators reach values up to $K \simeq 2$, thus confirming that correlators can be larger than 1.

Figure~\ref{fig:fig3}(b) shows the correlator difference $\Delta K(\tau) \equiv K_+(\tau) - K_-(\tau)$. This difference is more immune to offset fluctuations of the detector outputs, so the experimental $\Delta K(\tau)$  is less noisy than $K_\pm (\tau)$ in Fig.~\ref{fig:fig3}(a). The  experimental result (circles) in Fig.\ \ref{fig:fig3}(b)  agrees very well with the theoretical result (solid line)
$\Delta K(\tau) = \tan\varphi_{\rm a} \times 2c \, (\Omega_{\rm R}/\tilde \Omega_{\rm R}) \sin(\tilde \Omega_{\rm R}\tau)\,e^{-\Gamma\tau/2}$.
%

The correlator difference $\Delta K(\tau)$ can be useful to set $\varphi_{\rm a}=0$ accurately in experiments that need to avoid phase backaction. At present this is typically done by maximizing the response $\Delta I(\varphi_{\rm a})$, which depends quadratically on $\varphi_{\rm a}=0$ near the maximum, therefore leading to an inaccurate calibration. In contrast, $\Delta K(\tau)\propto \tan \varphi_{\rm a}$ vanishes at $\varphi_{\rm a}=0$ and depends linearly on $\varphi_{\rm a}$ in the vicinity (this still holds for the unscaled correlators), thus potentially providing a much better calibration accuracy. The practical use of $\Delta K(\tau)$ for this purpose needs further investigation.

{\it The GCR for multi-time correlators.} In the case of simultaneous CQM of $N_{\rm d}$ noncommuting qubit observables $\sigma_\ell\equiv\nb_\ell\boldsymbol{\sigma}$ (here $\boldsymbol{\sigma}$ is the vector of Pauli matrices, $\nb_\ell$ is the $\ell$th measurement axis direction on the Bloch sphere, and $\ell=1,2,...N_{\rm d}$), the GCR for an $N$-time correlator of the output signals $I_{\ell}(t)$ can be naturally generalized as  [cf.\ Eq.~\eqref{eq:coll-recipe}]
\begin{eqnarray}
\label{eq:coll-recipe-N}
&&\hspace{-0.0cm} K_{\ell_1...\ell_N}(t_1,...t_N) \equiv\, \langle I_{\ell_N}(t_N)\cdots I_{\ell_2}(t_2)\,I_{\ell_1}(t_1)\rangle
    \nonumber \\
&&\hspace{-0.0cm} = \!\!\! \sum_{\{I_{\ell_j}=\pm1\}}^{2^N} \bigg[  \prod_{j=2}^{j=N} I_{\ell_j}p\big(I_{\ell_j},t_j\big|I_{\ell_{j-1}},t_{j-1}\big) \bigg] I_{\ell_1}p\big(I_{\ell_1},t_1\big), \qquad
\end{eqnarray}
where the time arguments are ordered as $t_1<t_2<...<t_N$, $p\big(I_{\ell_1},t_1\big)$ is given by Eq.~\eqref{eq:pI1} with $\zb$ replaced by $\nb_{\ell_1}$, and the ``conditional probability'' factors are
\begin{equation}
\label{eq:cond-prob-N}
p\big(I_{\ell_j},t_j\big|I_{\ell_{j'}},t_{j'}\big) = \frac{1+I_{\ell_j}\nb_{\ell_j}\rb_{\rm ens}\big(t_{j}\big|I_{\ell_{j'}}\rb^{(j')}_{\rm coll},t_{j'}\big)}{2}.
\end{equation}
%
The collapsed state at time $t_{j}$ is $I_{\ell_j}\rb^{(j)}_{\rm coll}$, where
\begin{equation}
\label{eq:rcoll-N}
\rb^{(j)}_{\rm coll} = \nb_{\ell_{j}} + \mathcal{K}_{\ell_{j}}\nb_{\ell_{j}} \times \rb_{\rm ens}\big(t_{j}\big|I_{\ell_{j-1}}\rb^{(j-1)}_{\rm coll},t_{j-1}\big)
\end{equation}
for $j\geq 2$ [cf. Eq.~\eqref{eq:rcoll}]
and $\rb^{(1)}_{\rm coll}$ is given by Eq.~\eqref{eq:rcoll} with $\zb$ and $\mathcal{K}$ replaced by $\nb_{\ell_1}$ and $\mathcal{K}_{\ell_1}$, respectively. Parameters $\mathcal{K}_{\ell}=\tan \varphi_{\ell}^{\rm a}$ characterize the relative strength of phase backaction in the $\ell$th detector \cite{Shay2016}. In Eqs.~\eqref{eq:cond-prob-N}--\eqref{eq:rcoll-N}, $\rb_{\rm ens}$ obeys the evolution equation~\eqref{eq:ens-avg-evol}, where $\Lambda_{\rm ens}$ accounts for measurement of all $\sigma_\ell$.
This method to calculate $N$-time correlators is proven in the SM~\cite{SM}. Multi-time and/or multi-detector correlators can also exceed unity in the presence of phase backaction (with the coherent evolution not always needed) \cite{SM}.

{\it Conclusions.} We have developed a recipe for the calculation of correlators in continuous qubit measurements with phase backaction. As a consequence of the effective evolution outside the Bloch sphere, the normalized correlators can exceed 1. This has been confirmed experimentally, with the correlator reaching the value of 2. The correlators can be used as a calibration tool.

{\it Acknowledgements.} The work was supported by ARO grants  W911NF-15-1-0496 and W911NF-18-10178.

\newpage

\clearpage 
\onecolumngrid 
\vspace{\columnsep}
\begin{center}
\textbf{\large Supplemental Material for} \\
\vspace{0.2cm}
 \textbf{\large ``Correlators exceeding 1 in continuous measurements of superconducting qubits''}
\end{center}
\vspace{\columnsep}
\twocolumngrid

\setcounter{equation}{0}
\setcounter{figure}{0}
\setcounter{table}{0}
\setcounter{page}{1}

\renewcommand{\theequation}{S\arabic{equation}}
\renewcommand{\thefigure}{S\arabic{figure}}
\renewcommand{\bibnumfmt}[1]{[S#1]}
\renewcommand{\citenumfont}[1]{S#1}

\section{Experimental details}

\subsection{Setup and parameters}

We have performed continuous quantum measurement of the qubit observable $\sigma_z$ using the typical circuit QED setup, illustrated in Fig.~\ref{fig:setup} of the main text, generally similar to Ref.~\cite{Suppl-Murch2013} (though with important modifications). We use a 3D microwave cavity whose fundamental mode is dispersively coupled to a transmon qubit. The weakly-coupled input port is used to inject the Rabi drive and the readout tone. The stronger-coupled output port is used for the outgoing field. An additional cancellation tone (injected through circulator) displaces the outgoing field close to the vacuum, thus preventing saturation of the amplifier (the saturation becomes a serious problem for large angles $\varphi_{\rm a}$).

The cavity frequency is $6.66$~GHz and the qubit frequency is $4.26$~GHz (the same as in Refs.\ \cite{Suppl-Shay2016, Suppl-Atalaya2018npj}). The cavity mode decays with the rate $\kappa/2\pi =7.2$~MHz, the qubit relaxation times are $T_1=60\,\mu$s and $T_2^*=30 \, \mu$s. For qubit measurement, the cavity is coherently driven, causing the measurement-induced ensemble dephasing, which greatly exceeds intrinsic qubit dephasing. The resulting ensemble dephasing rate is $\Gamma =1/1.8\,\mu{\rm s}=2\pi\times 88$~kHz (for the results presented below in Sec.\ \ref{sec:other-angles}, $\Gamma =1/1.6\,\mu{\rm s}$).
The amplifier half-bandwidth is $B_{\rm amp}/2\pi \simeq 10$ MHz.
The detection quantum efficiency is $\eta = 0.44$.

For measurement of correlators, the qubit is prepared in the states $x_0=\pm 1$, and then we apply the Rabi rotation about $x$-axis with frequency $\Omega_{\rm R}/2\pi =\pm 1$ MHz (there are four combinations). The output signals from the continuous measurement are recorded for the duration of $4.88\,\mu$s with a timestep of 4~ns; after an additional averaging, the timestep is increased to $\Delta t=40$~ns. We use only the traces, selected by heralding the ground state of the qubit at the start of a run and checking that the transmon qubit is still within the two-level subspace after the run \cite{Suppl-Atalaya2018npj} (this eliminates about 25\% of traces).

Experimental parameters satisfy the relation $\Gamma \ll |\Omega_{\rm r}| \ll \kappa \alt B_{\rm amp}$. This justifies the white noise and the ``bad cavity'' assumptions needed for the quantum Bayesian formalism \cite{Suppl-Korotkov2011, Suppl-Korotkov2016}. Since $1/2 T_1\Gamma =0.015\ll 1$, we can neglect energy relaxation in the analysis.

\subsection{Calibration of response}
\label{sec:Calibration}

The response $\Delta I(\varphi_{\rm a})$ is calibrated for each angle $\varphi_{\rm a}$ between the amplified quadrature and the informational (maximum response) quadrature. For this calibration, the qubit is initialized in the state $|1\rangle$  ($z_{\rm in}=1$) or $|0\rangle$ ($z_{\rm in}=-1$) and then continuously measured with no Rabi oscillations applied. For each initial state, we collect about 17,000 traces of the continuous (digitized with $\Delta t$) output signal $\tilde I(t)$, each of 4 $\mu$s duration. Units of $\tilde I(t)$ are arbitrary, but always the same (same gain of the amplifier).

\begin{figure}[t!]
\centering
\begin{tabular}{c}
\includegraphics[width=0.95\linewidth, trim =1cm 0.5cm 1cm 0.9cm, clip=true]{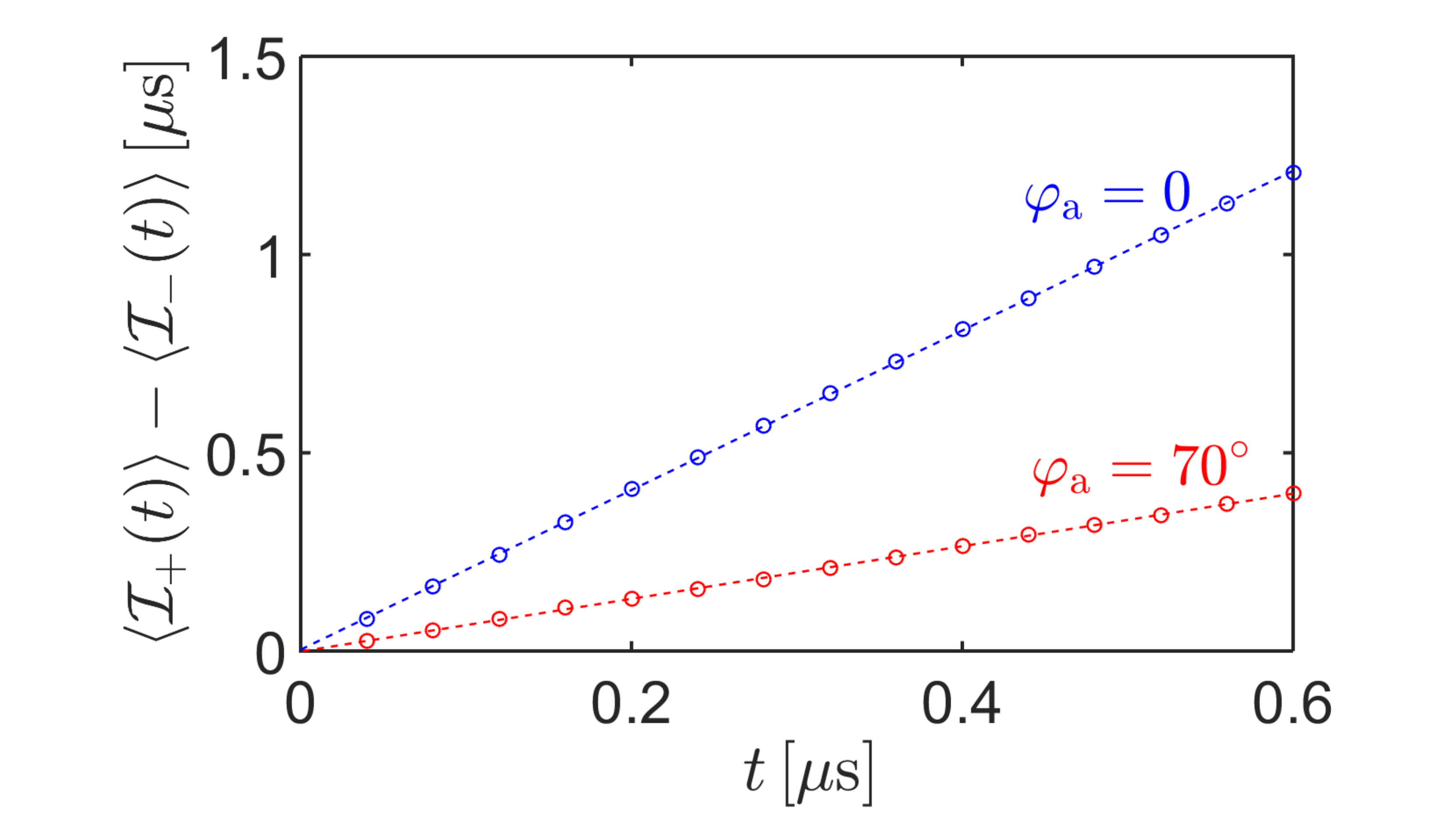}
\end{tabular}
\caption{Calibration of the detector response $\Delta I (\varphi_{\rm a})$ for $\varphi_{\rm a}=0$ and $70^\circ$. Detector response is obtained as the slope of the linear fit (dashed lines) to experimental results for $\langle \mathcal{I}_+(t)\rangle  - \langle \mathcal{I}_-(t)\rangle$, depicted by circles. We find $\Delta I (0) = 2.01$ and $\Delta I (70^\circ)=0.66$.  }
\label{fig:Suppl-fig1}
\end{figure}

To find the response $\Delta I(\varphi_{\rm a})$, for each trace we numerically calculate the integral
\begin{align}
\label{eq:int-I}
\mathcal{I}_\pm(t) = \int_0^t \tilde I_\pm(t')\, dt',
\end{align}
where the subscript $\pm$ corresponds to initial state $z_{\rm in}=\pm 1$, and then average over the ensemble of traces to get $\langle \mathcal{I}_\pm (t)\rangle$. The difference $\langle \mathcal{I}_+ (t)\rangle -\langle \mathcal{I}_- (t)\rangle$ for $\varphi_{\rm a}=0$ and $\varphi_{\rm a}=70^\circ$ is shown in Fig.\ \ref{fig:Suppl-fig1}. From the slope of these practically straight lines, we find the response $\Delta I(\varphi_{\rm a}) = d[\langle \mathcal{I}_+(t)\rangle - \langle \mathcal{I}_-(t) \rangle]/dt$. Note that we use only initial 0.6 $\mu$s of the process, because for a significantly longer integration there is a noticeable deviation from straight lines due to energy relaxation. From the slopes of lines in Fig.\ \ref{fig:Suppl-fig1}, we obtain the responses $\Delta I (0)=\Delta I_{\rm max}=2.01$ and  $\Delta I (70^\circ )=0.66$. This confirms the expected relation $\Delta I(\varphi_{\rm a}) = \Delta I_{\max}\cos\varphi_{\rm a}$ within 3\% inaccuracy.

\begin{figure}[t!]
\centering
\begin{tabular}{c}
\includegraphics[width=0.95\linewidth, trim =1cm 0.5cm 1cm 1cm, clip=true]{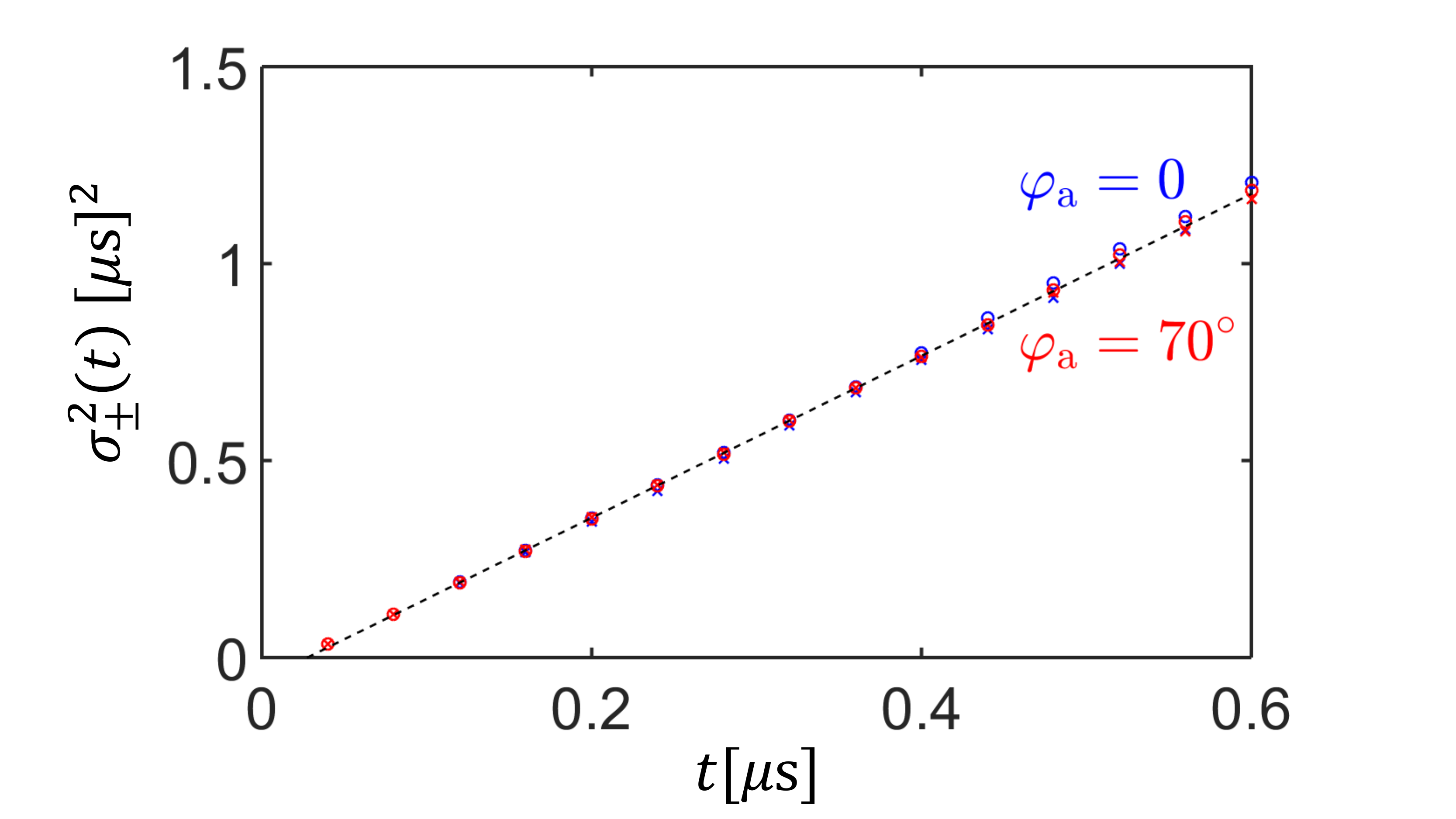}
\end{tabular}
\caption{The variance $\sigma^2_\pm(t)\equiv \langle \mathcal{I}_\pm^2(t)\rangle - \langle \mathcal{I}_\pm(t)\rangle^2$ as a function of the integration time $t$. Circles show $\sigma_+$, crosses show $\sigma_-$, blue symbols are for $\varphi_{\rm a}=0$, red symbols are for $\varphi_{\rm a}=70^\circ$. All four cases can be fitted by one straight (dashed) line with slope $d\sigma^2(t)/dt= 2.06 \,\mu$s, which gives $\tau_{\rm m}(0)=2.04 \,\mu$s and $\tau_{\rm m}(70^\circ)=18.9 \,\mu$s.  }
\label{fig:Suppl-fig2}
\end{figure}

To find the quantum efficiency $\eta$ (even though we do not actually need it for the correlators), we first obtain the ``measurement time'' $\tau_{\rm m}$ as  $\tau_{\rm m} (\varphi_{\rm a})= [2/\Delta I(\varphi_{\rm a})]^2\times d\sigma^2(t)/dt$, where the variance $ \sigma^2(t) =\sigma^2_\pm(t) \equiv \langle \mathcal{I}_\pm^2(t)\rangle - \langle \mathcal{I}_\pm(t)\rangle^2$ should theoretically be independent of $\varphi_{\rm a}$ and $z_{\rm in}$.
Figure~\ref{fig:Suppl-fig2} shows that indeed $\sigma^2_+(t)\approx \sigma^2_-(t)$, and they are almost the same for $\varphi_{\rm a}=0$ and $\varphi_{\rm a}=70^\circ$, so we practically have one straight line. From the linear fit, $d\sigma^2(t)/dt=2.06\, \mu{\rm s}$, we obtain $\tau_{\rm m}(0)=\tau_{\min}\approx 2.04\,\mu$s and  $\tau_{\rm m}(70^\circ) = 18.9\,\mu$s. Therefore, the quantum efficiency is  $\eta = (2\Gamma\tau_{\min})^{-1} = 0.44$.

\subsection{Correlators}
\label{sec:expt-corr-calculation}

For measurement of correlators, the qubit is prepared at time $t_0=0$ in the pure state $\rb_{\rm 0}=(\pm 1,0,0)$ and then is Rabi-rotated about $x$-axis with frequency $\Omega_{\rm R}/2\pi =\pm 1$~MHz (four combinations), while being continuously measured along $z$-axis. The ensemble-averaged evolution is supposed to change (decrease) only $x$ component of the qubit state, while $z$ and $y$ components should remain zero on average. We obtain experimental correlators as
\begin{align}
\label{eq:Suppl-K-expt}
& K (\tau) = \frac{1}{T} \int_{t_{\rm skip}}^{t_{\rm skip+T}} \bigg\langle \frac{ \tilde I(t_1) - \langle \tilde I(t_1)\rangle}{\Delta I(\varphi_{\rm a})}
    \nonumber \\
&\hspace{2.5cm} \times  \frac{ \tilde I(t_1+\tau) - \langle \tilde I(t_1+\tau)\rangle}{\Delta I(\varphi_{\rm a})} \bigg\rangle \, dt_1,
\end{align}
where the averaging time is $T=0.28\,\mu$s (to reduce fluctuations) and the discarded initial duration is $t_{\rm skip}=0.28\,\mu$s (to avoid initial transients in the data). Note that both $T$ and $t_{\rm skip}$ are small in comparison with $1/\Gamma=1.8\, \mu$s and duration of 4.88 $\mu$s of the recorded traces.

\begin{figure}[t!]
\centering
\begin{tabular}{c}
\includegraphics[width=0.95\linewidth, trim =0cm 0.5cm 2cm 0.5cm, clip=true]{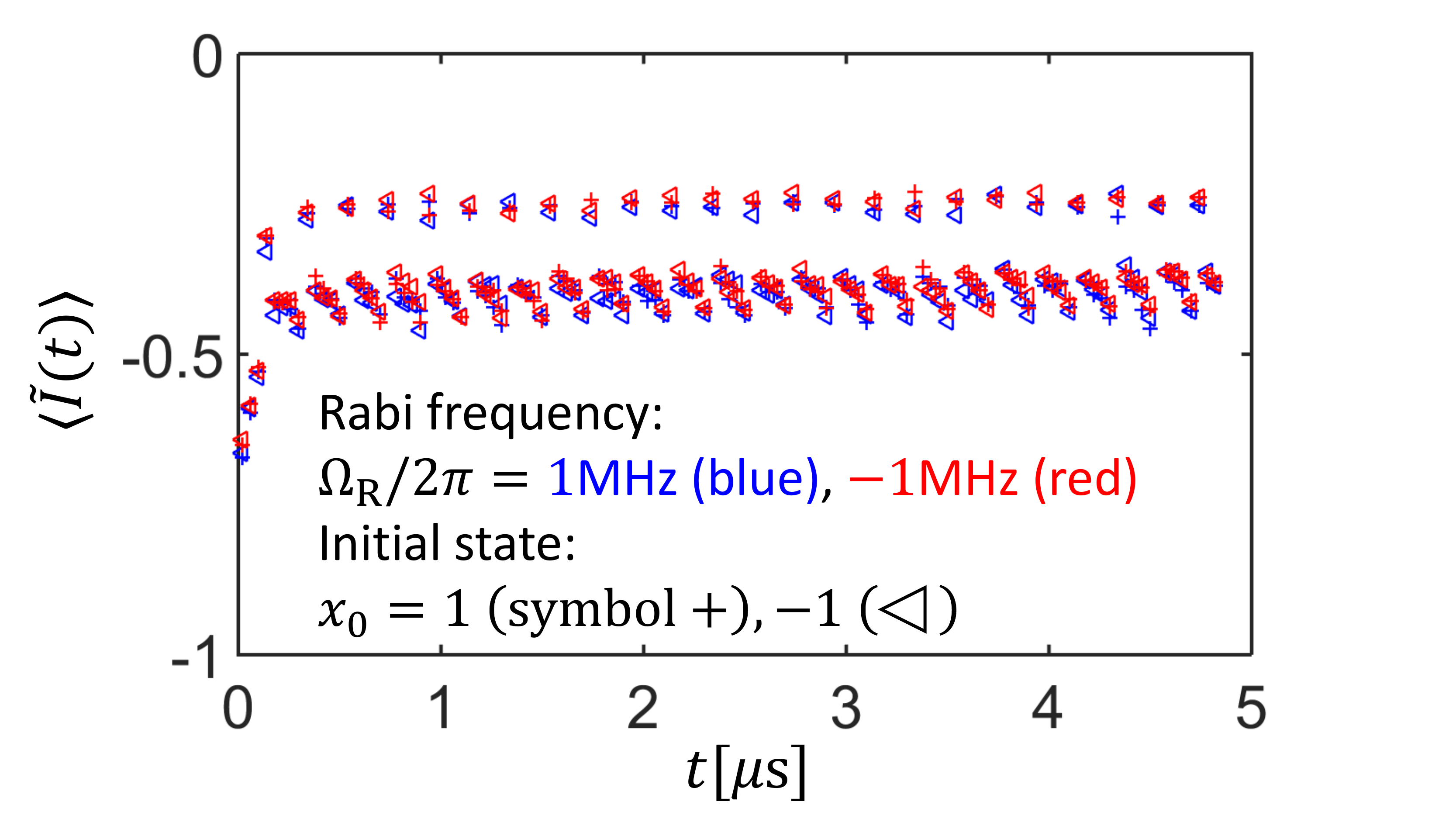}
\end{tabular}
\caption{The offset $\langle \tilde I(t)\rangle$ for the initial state $x_0=1$ (crosses) or $x_0=-1$ (triangles) and Rabi frequency $\Omega_{\rm R}= 1$ MHz (blue symbols) or $-1$ MHz (red symbols). The data points are separated by $\Delta t=40$ ns. }
\label{fig:Suppl-fig-I}
\end{figure}

Since on average $z(t)=0$ for $x_0=\pm 1$ and Rabi rotation over $x$-axis, the average $\langle \tilde I(t)\rangle$ in Eq.\ (\ref{eq:Suppl-K-expt}) should theoretically be a constant offset $\tilde I_{\rm o}$. However, this is not exactly the case in the experiment, as seen from Fig.\ \ref{fig:Suppl-fig-I}, which shows $\langle \tilde I(t)\rangle$ for all four combinations of $x_0$ and $\Omega_{\rm R}$ in the case $\varphi_{\rm a}=70^\circ$. Besides the overall shift, $\tilde I_{\rm o}\simeq -0.4$, we see small periodic features, the reason for which is unclear. Note that the size of these features ($\simeq \pm 0.1$) is small in comparison with the response (0.66) and noise in an individual trace ($\sigma_{\Delta t}\approx 6$); however, they still slightly affect the correlators. This is why we subtract $\langle \tilde I(t)\rangle$ in Eq.\ (\ref{eq:Suppl-K-expt}) instead of subtracting a constant offset $\tilde I_o$, in order to remove the fluctuating offsets. Moreover, we calculate $\langle \tilde I(t)\rangle$ in Eq.\ (\ref{eq:Suppl-K-expt}) by averaging over a relatively small number of neighboring runs (about 3,000), in order to account for offsets, slowly fluctuating in time. Figure \ref{fig:Suppl-fig-I} also explains why we use $t_{\rm skip}= 0.28 \, \mu$s, i.e., skip first seven data points, for which some transient process is easily noticeable.

\vspace{0.3cm}

To calculate the theoretical result for the two-time correlator, we use Eq.\ (20) of the main text with $\rb(t_1-0)=(e^{-\Gamma t_1}x_0,0,0)$. Solving the ensemble-averaged qubit evolution (energy relaxation is neglected), we obtain the correlator
\begin{eqnarray}
\label{eq:Suppl-integrand}
&& K(t_1,t_1+\tau) = \left [\cos (\tilde\Omega_{\rm R}\tau) + \frac{\Gamma}{2\tilde\Omega_{\rm R}}\sin(\tilde\Omega_{\rm R} \tau)\right] e^{-\Gamma\tau/2}
    \nonumber \\
&& \hspace{2.1cm} + x_{\rm 0}\,e^{-\Gamma t_1}\,\frac{\tan\varphi_{\rm a}\,\Omega_{\rm R}}{\tilde\Omega_{\rm R}}\sin(\tilde\Omega_{\rm R} \tau)\, e^{-\Gamma \tau/2}, \qquad
\end{eqnarray}
where $\tilde\Omega_{\rm R} = \sqrt{\Omega_{\rm R}^2 - \Gamma^2/4}$. To perform the additional integration over $t_1$ in Eq.\ \eqref{eq:Suppl-K-expt}, we notice that $t_1$ enters Eq.\ \eqref{eq:Suppl-integrand} only via the factor $e^{-\Gamma t_1}$ in the second term. Therefore, the only change in Eq.\ \eqref{eq:Suppl-integrand} is the replacement
    \begin{equation}
x_0\to c \, x_0 ,\,\,\, c = e^{-\Gamma t_{\rm skip}}\,
\frac{1-e^{-\Gamma T}}{\Gamma T}.
    \end{equation}
Thus we obtain Eq.\ (23) of the main text,
\begin{eqnarray}
\label{eq:Suppl-K(tau)-an}
&& K(\tau) = \left [\cos (\tilde\Omega_{\rm R}\tau) + \frac{\Gamma}{2\tilde\Omega_{\rm R}}\sin(\tilde\Omega_{\rm R} \tau)\right] e^{-\Gamma\tau/2}
    \nonumber \\
&& \hspace{1.2cm} + c\, x_{\rm 0}\, \tan\varphi_{\rm a}\,\frac{\Omega_{\rm R}}{\tilde\Omega_{\rm R}}\sin(\tilde\Omega_{\rm R} \tau)\, e^{-\Gamma \tau/2}.
\end{eqnarray}

\begin{figure}[t!]
\centering
\begin{tabular}{c}
\includegraphics[width=0.9\linewidth, trim =0cm 3.4cm 0cm 0cm, clip=true]{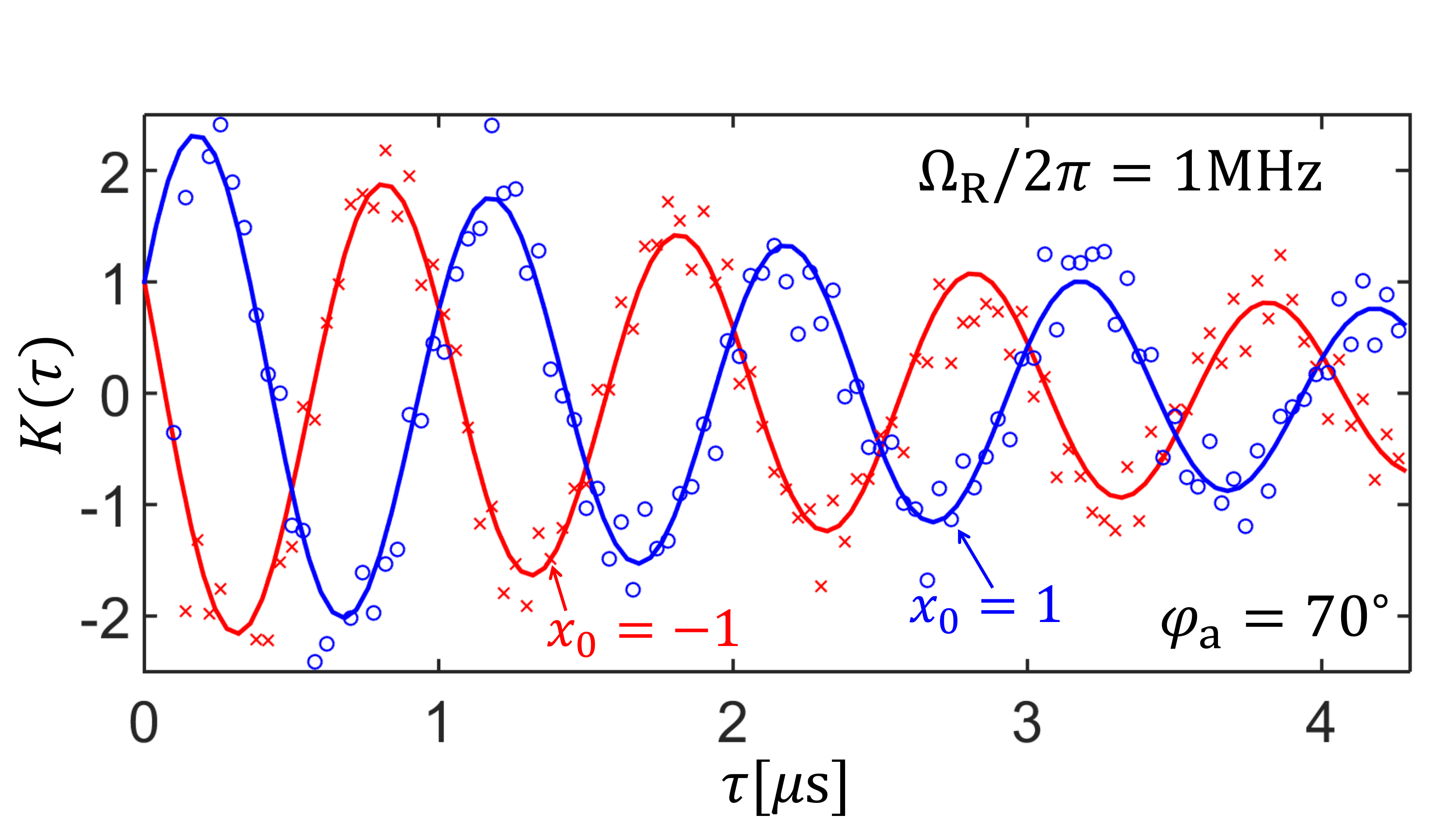} \\
\includegraphics[width=0.9\linewidth, trim =0cm 0cm 0cm 1.cm, clip=true]{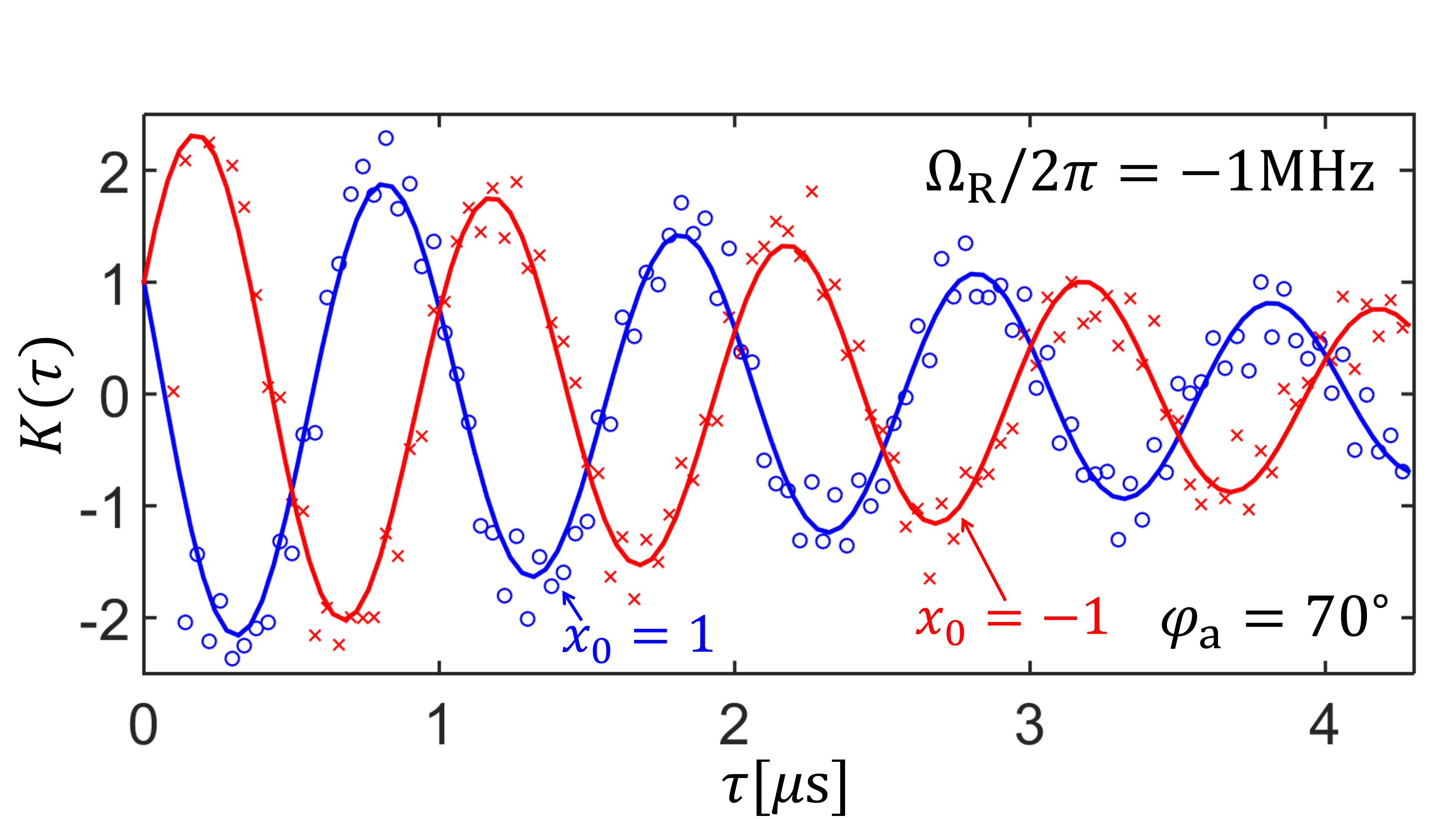}
\end{tabular}
\caption{Experimental correlators (symbols) and analytics (lines) for the four cases with $\Omega_{\rm R}/2\pi=\pm 1$~MHz and $x_{\rm 0}=\pm 1$. The amplified-quadrature angle determining the phase backaction is $\varphi_{\rm a}=70^\circ$, time-averaging parameters are $T=0.28\,\mu$s and $t_{\rm skip}=0.28 \, \mu$s, ensemble averaging is over $3.2\times 10^5$ traces in each case.  }
\label{fig:Suppl-fig3}
\end{figure}

Figure \ref{fig:Suppl-fig3} shows experimental results (symbols) and analytics (lines) for the correlators $K(\tau)$ for $\varphi_{\rm a}=70^\circ$ in the four cases: for Rabi frequency $\Omega_{\rm R}/2\pi=1$ MHz (upper panel) or $-1$ MHz (lower panel) and initial state $x_0=1$ (blue circles and blue lines) or $x_0=-1$ (red crosses and red lines). There is a good agreement between the theory and experiment in all the four cases.
In Fig.\ 3(a) of the main text we present the same results, additionally averaged over two cases with the same product $\Omega_{\rm R}x_0$.

\subsection{Correlators for other angles $\varphi_{\rm a}$}
\label{sec:other-angles}

We have also measured the correlators for angles $\varphi_{\rm a}=0$, $40^\circ$, and $80^\circ$. This was done on a different date compared with the results presented in Sections \ref{sec:Calibration}, \ref{sec:expt-corr-calculation}, and in the main text, so parameters are slightly different. In particular, the qubit ensemble dephasing rate during measurement is $\Gamma =1/1.6\, \mu$s (a slightly higher microwave power for measurement). The detector responses are $\Delta I(0)=2.3$, $\Delta I(40^\circ)=1.75$, and $\Delta I(80^\circ)=0.44$. The relation
$\Delta I(\varphi_{\rm a}) = \Delta I_{\rm max}\cos(\varphi_{\rm a})$ is satisfied with 1\% inaccuracy for $40^\circ$ and with 10\% inaccuracy for $80^\circ$ (inaccuracy grows with decrease of the SNR).

\begin{figure}[t!]
\centering
\begin{tabular}{c}
\includegraphics[width=0.9\linewidth, trim =0cm 3.4cm 0cm 0cm, clip=true]{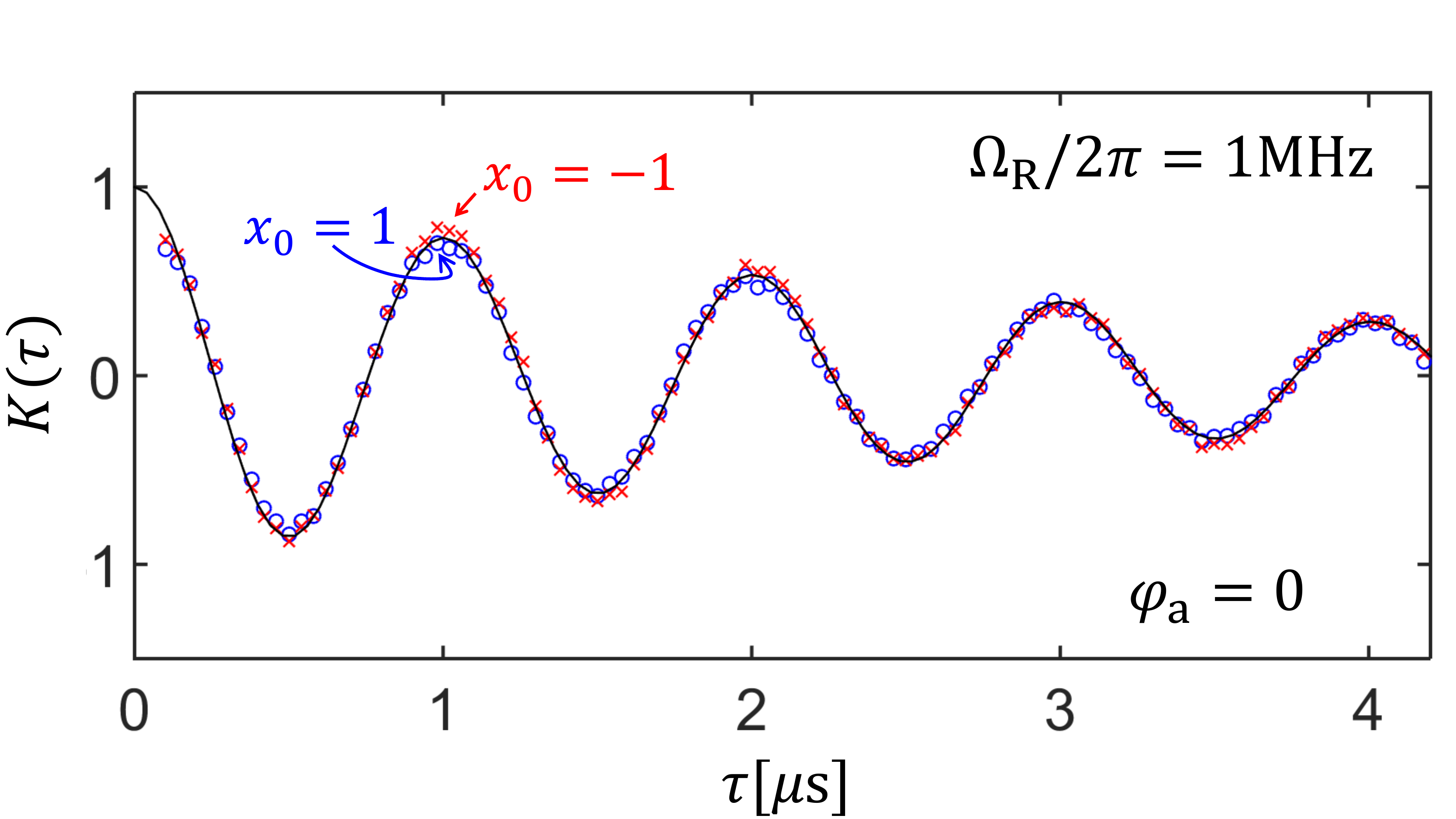} \\
\includegraphics[width=0.9\linewidth, trim =0cm 3.4cm 0cm 1.cm, clip=true]{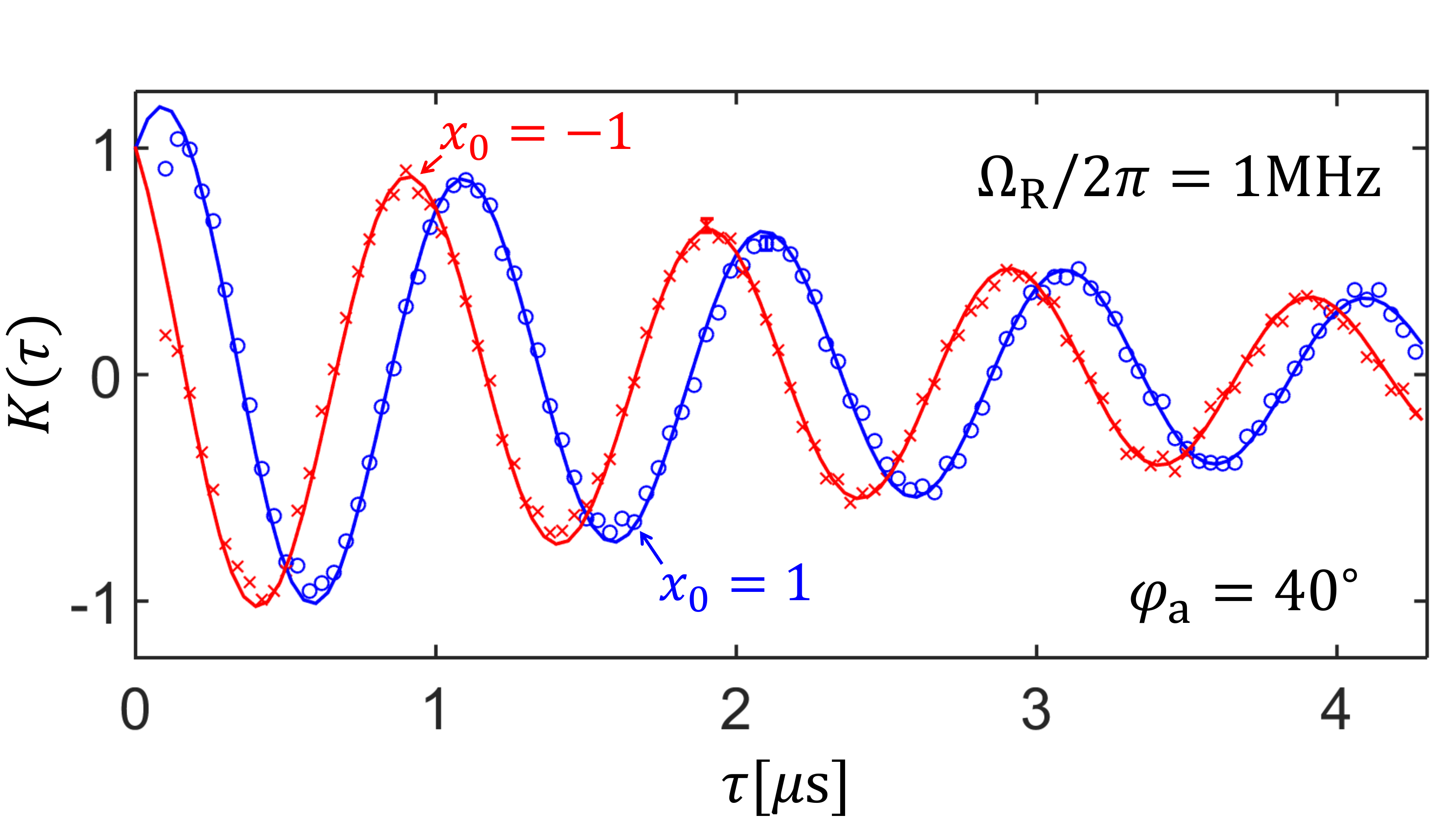} \\
\includegraphics[width=0.9\linewidth, trim =0cm 0cm 0cm 1.cm, clip=true]{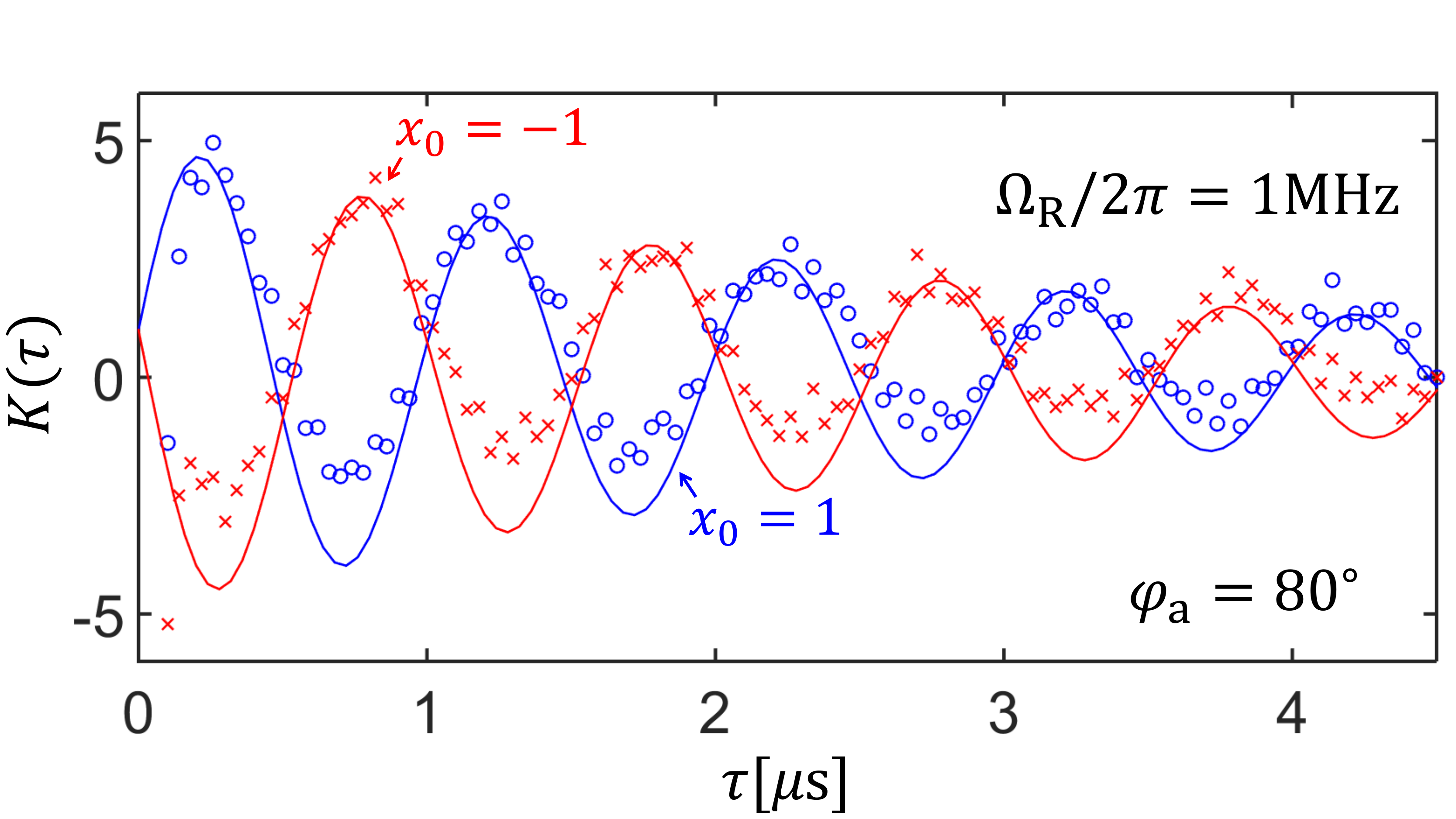}
\end{tabular}
\caption{Experimental correlators (symbols) and theoretical predictions (lines) for angles $\varphi_{\rm a}=0$ (top panel), $40^\circ$ (middle panel), and $80^\circ$ (bottom panel). Initial states are $x_0=1$ (blue circles and lines) and $x_0=-1$ (red crosses and lines), Rabi frequency is $\Omega_{\rm R}=1$ MHz.   }
\label{fig:Suppl-fig4}
\end{figure}

Figure~\ref{fig:Suppl-fig4} shows the experimental correlators (symbols) and theoretical results (lines) for the angles $\varphi_{\rm a}=0,40^\circ$ and $80^\circ$. We use $\Omega_{\rm R}/2\pi = 1$ MHz (only one direction) and $x_0=\pm 1$, the time-integration parameters are still $T=t_{\rm skip}=0.28\, \mu$s.  The experimental correlators for $\varphi_{\rm a}=0$ agree with the theory very well; they are practically the same for $x_0=1$ and $x_0=-1$ (theoretically there is no dependence on the initial state \cite{Suppl-Atalaya2018npj}), and $|K(\tau)|\leq 1$ always because there is no phase backaction. Experimental correlators for $\varphi_{\rm a}=40^\circ$ also agree well with the theory; the correlator $K(\tau)$ for $x_0=1$ marginally exceeds 1 at only one point. Experimental correlators  for $\varphi_{\rm a}=80^\circ$ greatly exceed 1 at many points, reaching values up to $K_{\rm max}\simeq 5$. However, there is a significant deviation from the theory, which is somewhat expected since the SNR greatly decreases for angles $\varphi_{\rm a}$ close to $\pi/2$.

\section{Generalized collapse recipe for multi-time multi-detector correlators}

In this section we prove the generalized collapse recipe (GCR) for multi-time correlators from simultaneous continuous measurement of $N_{\rm d}$ noncommuting qubit observables $\sigma_\ell=\nb_\ell\boldsymbol{\sigma}$, where  $\nb_\ell$ is the $\ell$th measurement axis direction on the Bloch sphere and $\ell=1,...N_{\rm d}$.

In this case, the quantum Bayesian equation for qubit evolution in It\^o interpretation is [cf.\ Eq.~\eqref{eq:QBE} of the main text]
\begin{equation}
\label{eq:Suppl-QBE}
\dot\rb = \Lambda_{\rm ens}(\rb -\rb_{\rm st})+ \sum_{\ell=1}^{N_{\rm d}} \left[\frac{\nb_\ell- (\nb_\ell \rb)\, \rb}{\sqrt{\tau_{\ell}}} +\mathcal{K}_\ell\,\frac{(\nb_\ell\boldsymbol \! \times \! \rb)}{\sqrt{\tau_{\ell}}}\right]\xi_\ell(t),
\end{equation}
where $\tau_\ell$ is the ``measurement time'' for the $\ell$th detector and
$\mathcal{K}_\ell=\tan \varphi^{\rm a}_\ell$ determines the corresponding relative strength of phase backaction. The normalized output signal from the $\ell$th detector is modeled as
\begin{equation}
\label{eq:Suppl-I-ell}
I_\ell(t) = {\rm Tr}[\sigma_\ell\rho(t)] + \sqrt{\tau_\ell}\,\xi_\ell(t) = \nb_\ell\rb(t) + \sqrt{\tau_\ell}\,\xi_\ell(t),
\end{equation}
where $\xi_\ell$ are uncorrelated white noises,
\begin{equation}
\label{eq:Suppl-xi-corr}
\langle \xi_\ell(t) \, \xi_{\ell'}(t')\rangle = \delta_{\ell \ell'}\, \delta(t-t').
\end{equation}

Let us consider the $N$-time correlator
\begin{equation}
\label{eq:Suppl-KN-def}
K_{\ell_1...\ell_N}(t_1,...t_N) \equiv \langle I_{\ell_N}(t_N) \cdots I_{\ell_2}(t_2)\, I_{\ell_1}(t_1)\rangle ,
\end{equation}
in which  the time arguments are ordered as $t_1<t_2<...<t_N$ and $N$ can be smaller, equal, or larger than $N_{\rm d}$.
We will prove that this correlator can be obtained from the GCR formula
\begin{eqnarray}
\label{eq:Suppl-coll-recipe-N}
&&\hspace{-0.0cm}K^{\rm GCR}_{\ell_1...\ell_N}(t_1,...t_N)
    \nonumber \\
&& \hspace{0.3cm} =\!\!\!  \sum_{\{I_{\ell_j} =\pm1\}}^{2^N} \bigg[ \prod_{j=2}^{j=N} I_{\ell_j}p\big(I_{\ell_j},t_j\big|I_{\ell_{j-1}},t_{j-1}\big)\bigg] I_{\ell_1}p\big(I_{\ell_1},t_1\big) ,
    \nonumber\\
\end{eqnarray}
where the sum is over $2^N$ scenarios of obtaining discrete outcomes $I_{\ell_j}=\pm 1$ of (fictitious) ``strong'' measurements at time moments $t_j$ ($j=1,...N$),
\begin{equation}
\label{eq:Suppl-pI1}
p\big(I_{\ell_1},t_1\big) = \frac{1 + I_{\ell_1}\, \nb_{\ell_1}\rb (t_1 -0) }{2}
\end{equation}
is the probability to get the first outcome $I_{\ell_1}=\pm 1$ at time $t_1$, and
\begin{equation}
\label{eq:Suppl-cond-prob-j}
p\big(I_{\ell_j},t_j\big|I_{\ell_{j-1}},t_{j-1}\big) = \frac{1+I_{\ell_j}\nb_{\ell_j}\rb_{\rm ens}\big(t_{j}\big|I_{\ell_{j-1}}\rb^{(j-1)}_{\rm coll},t_{j-1}\big)}{2}
\end{equation}
is the ``conditional probability'' to get the outcome $I_{\ell_j}$ at time $t_j$ ($j\geq 2$) given that we got the outcome $I_{\ell_{j-1}}$ at time $t_{j-1}$ (this ``probability'' can be negative or larger than 1).  We assume (pretend) that the strong measurement of $\sigma_{\ell_j}$ (with phase backaction) at time $t_j$ with the result $I_{\ell_j}=\pm 1$ collapses (abruptly moves) the qubit state to
    \begin{equation}
\rb (t_j+0)= I_{\ell_j} \rb^{(j)}_{\rm coll}  = I_{\ell_j} \big [ \nb_{\ell_{j}} + \mathcal{K}_{\ell_{j}}\, \nb_{\ell_{j}} \times \rb (t_j-0)  \big] ,
    \label{eq:Suppl-coll-state}\end{equation}
while at other times, $t\neq t_j$, the qubit evolution is given by the ensemble-averaged equation
\begin{equation}
\label{eq:Suppl-ens-avg-evol}
\dot\rb_{\rm ens} = \Lambda_{\rm ens}(\rb_{\rm ens} - \rb_{\rm st}).
\end{equation}
Therefore, in each of the $2^N$ scenarios, we have a different sequence of after-collapse states $I_{\ell_j} \rb^{(j)}_{\rm coll}$, with
\begin{equation}
\label{eq:Suppl-rcoll-1}
\rb^{(1)}_{\rm coll} = \nb_{\ell_1} + \mathcal{K}_{\ell_1}\, \nb_{\ell_1} \times \rb (t_1-0)
\end{equation}
for the first collapse, and then for $j\geq 2$ we have
\begin{equation}
\label{eq:Suppl-rcoll-j}
\rb^{(j)}_{\rm coll} = \nb_{\ell_{j}} + \mathcal{K}_{\ell_{j}}\nb_{\ell_{j}} \times \rb_{\rm ens}\big(t_{j}\big|I_{\ell_{j-1}}\rb^{(j-1)}_{\rm coll},t_{j-1}\big) ,
\end{equation}
where $\rb_{\rm ens}\big(t\big| \rb_{\rm in}, t_{\rm in}\big)$ is the solution of Eq.\ (\ref{eq:Suppl-ens-avg-evol}) with initial condition
$\rb_{\rm ens}\big(t_{\rm in}\big| \rb_{\rm in}, t_{\rm in}\big)=\rb_{\rm in}$, and $\rb (t_1-0)=\rb_{\rm ens}\big(t_1\big| \rb_{\rm 0}, t_{\rm 0}\big)$ if the procedure starts at time $t_0<t_1$ with the initial state $\rb_0$.

Note that the initial qubit state should be physical, and therefore the 3-vector $\rb_0$ should be within the Bloch sphere, $| \rb_0| \leq 1$. However, after each collapse, the state $I_{\ell_j} \rb^{(j)}_{\rm coll}$ will be outside the Bloch sphere (if $\mathcal{K}_{\ell_{j}}\neq 0$). Therefore, the state before the next collapse may also be outside the Bloch sphere, and then the ``conditional probabilities'' for the next outcome  $I_{\ell_{j+1}}=\pm 1$  may be negative or larger than 1  -- see Eq.\ (\ref{eq:Suppl-cond-prob-j}). Also note that the $3\times3$ matrix $\Lambda_{\rm ens}$ in Eq.~\eqref{eq:Suppl-ens-avg-evol} takes into account unitary evolution, continuous measurement by all $N_{\rm d}$ detectors, and possible additional decoherence. Both $\Lambda_{\rm ens}$ and $\rb_{\rm st}$ can depend on time. The formal solution of Eq.~\eqref{eq:Suppl-ens-avg-evol} can still be written in the same form as in the main text,
    \begin{equation}
\label{eq:Suppl-rens-sol}
\rb_{\rm ens}\big(t\big|\rb_{\rm in},t_{\rm in}\big) = \mathcal{P}(t |t_{\rm in})\, \rb_{\rm in} + \boldsymbol{\mathcal{P}}_{\rm st}(t|t_{\rm in}),
    \end{equation}
where $\mathcal{P}(t|t')$ is a $3\!\times\! 3$ matrix satisfying  equation $\partial_t \mathcal{P}(t|t') = \Lambda_{\rm ens} (t) \,\mathcal{P}(t|t')$ with $\mathcal{P}(t'|t')=\openone$, and  $\boldsymbol{\mathcal{P}}_{\rm st}(t|t') = -\int_{t'}^t  \mathcal{P}(t|t'')\, \Lambda_{\rm ens}(t'')\, \rb_{\rm st}(t'')\, dt''$.

\vspace{0.2cm}

To prove that Eqs.~\eqref{eq:Suppl-coll-recipe-N}--\eqref{eq:Suppl-rcoll-j} give the correct value for the multi-time correlator~\eqref{eq:Suppl-KN-def}, let us first carry out the summation over the last outcome $I_{\ell_N}$ in Eq.~\eqref{eq:Suppl-coll-recipe-N} and represent the result as
\begin{equation}
\label{eq:Suppl-KN-alternative-form-GCR}
K^{\rm GCR}_{\ell_1...\ell_N}(t_1,...t_N) = \nb_{\ell_N} \Kb^{\rm GCR}_{\ell_1...\ell_N}(t_1,...t_N),
\end{equation}
where we have introduced the vector-valued correlator
\begin{align}
\label{eq:Suppl-KN-vector-GCR}
\Kb^{\rm GCR}_{\ell_1...\ell_N}(t_1,...t_N)\equiv & \sum_{\{I_{\ell_j}=\pm1\}}^{2^{N-1}}  \rb_{\rm ens}\big(t_N\big|I_{\ell_{N-1}}\rb_{\rm coll}^{(N-1)},t_{N-1}\big)
    \nonumber \\
&\hspace{-1.7cm}\times \bigg[ \prod_{j=2}^{j=N-1} I_{\ell_j}p\big(I_{\ell_j},t_j\big|I_{\ell_{j-1}},t_{j-1}\big)\bigg] I_{\ell_1}p\big(I_{\ell_1},t_1\big).
\end{align}
We then apply Eq.~\eqref{eq:Suppl-rens-sol} to Eq.~\eqref{eq:Suppl-KN-vector-GCR}, use Eq.~\eqref{eq:Suppl-rcoll-j} with $j=N-1$ and use the relations~\eqref{eq:Suppl-KN-def} and \eqref{eq:Suppl-KN-alternative-form-GCR}--\eqref{eq:Suppl-KN-vector-GCR} to obtain the recursive formula
\begin{align}
\label{eq:Suppl-Kvec-N-recursive-1-GCR}
&\Kb^{\rm GCR}_{N} = \mathcal{P}(t_N|t_{N-1})\big[\nb_{\ell_{N-1}}K^{\rm GCR}_{N-2}
    \nonumber \\
&\hspace{0.7cm} +  \mathcal{K}_{\ell_{N-1}} \nb_{\ell_{N-1}} \times \Kb^{\rm GCR}_{N-1}\big]+
 K^{\rm GCR}_{N-1}\,\boldsymbol{\mathcal{P}}_{\rm st}(t_N|t_{N-1}),
\end{align}
where for brevity  $\Kb^{\rm GCR}_N\equiv \Kb^{\rm GCR}_{\ell_1...\ell_N}(t_1,...t_N)$ and $K^{\rm GCR}_N\equiv K^{\rm GCR}_{\ell_1...\ell_N}(t_1,...t_N)$. This recursion for $N$ needs two initial cases, for which $N=2$ and $N=1$ can be used. The correlators
for $N=1$ are trivial,
\begin{align}
& \Kb_{\ell_1}^{\rm GCR} (t_1)=\rb (t_1-0)
\label{eq:Suppl-recursive-2-GCR}
\end{align}
and therefore $K_{\ell_1}^{\rm GCR}(t_1)= \nb_{\ell_1}\rb (t_1-0)$,
while the GCR correlators for $N=2$ are [cf.\ Eq.\ (10) of the main text]
    \begin{eqnarray}
&& \Kb^{\rm GCR}_{\ell_1 \ell_2} (t_1,t_2)=
 \rb_{\rm ens} \big( t_2 \big|\rb_{\rm coll}^{(1)},t_1 \big) \, \frac{1+\nb_{\ell_1}\rb (t_1-0)}{2}
 \nonumber \\
&& \hspace{1.52cm}
 -  \rb_{\rm ens} \big( t_2 \big| -\rb_{\rm coll}^{(1)}, t_1 \big)
  \frac{1-\nb_{\ell_1}\rb (t_1-0)}{2}  \qquad
    \label{eq:Suppl-Kb2}\end{eqnarray}
and correspondingly $K^{\rm GCR}_{\ell_1 \ell_2} (t_1,t_2) = \nb_{\ell_2} \Kb^{\rm GCR}_{\ell_1 \ell_2} (t_1,t_2)$. Using Eq.\ (\ref{eq:Suppl-rens-sol}), it is easy to see that Eq.\ (\ref{eq:Suppl-Kb2}) can be obtained from the recursion (\ref{eq:Suppl-Kvec-N-recursive-1-GCR}) if we formally define
    \begin{equation}
K_0^{\rm GCR}=1 .
    \label{eq:Suppl-K0}\end{equation}

Thus far, we have just rewritten the GCR in a recursive form [Eqs.~\eqref{eq:Suppl-KN-alternative-form-GCR} and \eqref{eq:Suppl-Kvec-N-recursive-1-GCR}]. Next, we will show that the same recursive relations for the correlators [including the initial cases \eqref{eq:Suppl-recursive-2-GCR}--(\ref{eq:Suppl-K0})]  can be obtained from the quantum Bayesian equations~Eq.~\eqref{eq:Suppl-QBE}--\eqref{eq:Suppl-I-ell}, thus proving the GCR.

Now we are considering the actual process (not the fictitious scenarios of the GCR), so $I_{\ell}(t)$ are continuous noisy signals -- see Eq.\ (\ref{eq:Suppl-I-ell}). Using the causality property $\langle \xi_{\ell}(t)\,I_{\ell'}(t')\rangle=0$ for $t>t'$, we can express the  multi-time correlator \eqref{eq:Suppl-KN-def} in the same form as Eq.~\eqref{eq:Suppl-KN-alternative-form-GCR},
\begin{align}
\label{eq:Suppl-KN}
K_{\ell_1...\ell_N}(t_1,...t_N) =\, \nb_{\ell_N}\Kb_{\ell_1...\ell_N} (t_1,...t_N),
\end{align}
where we have introduced the vector-valued correlator
\begin{align}
\label{eq:Suppl-KN-vector}
\Kb_{\ell_1...\ell_N} (t_1,...t_N) \equiv  \langle \rb_N \, I_{\ell_{N-1}}(t_{N-1})...I_{\ell_1}(t_1)\rangle
\end{align}
and for brevity we use notation $\rb_N\equiv\rb(t_{N})$.  Also introducing the short notation $\Kb_N\equiv \Kb_{\ell_1...\ell_N} (t_1,...t_N)$ and using Eq.~\eqref{eq:Suppl-I-ell} for $I_{\ell_{N-1}}(t)$, we can write $\Kb_N$ as a sum of two terms,
\begin{subequations}
\label{eq:Suppl-KN-decomposition}
\begin{align}
\label{eq:Suppl-KN-decomposition-0}
&\Kb_N = \Kb_N^{(1)} + \Kb_N^{(2)}, \\
&\Kb_N^{(1)} \equiv  \big\langle \rb_N \, \big(\nb_{\ell_{N-1}}\rb_{N-1}\big)\,I_{\ell_{N-2}}(t_{N-2})...I_{\ell_1}(t_1)\big\rangle, \label{eq:Suppl-KN-decomposition-1}\\
&\Kb_N^{(2)} \equiv  \big\langle  \rb_N \, \sqrt{\tau_{\ell_{N-1}}} \xi_{\ell_{N-1}}(t_{N-1})\,I_{\ell_{N-2}}(t_{N-2})...I_{\ell_1}(t_1)\big \rangle.   \label{eq:Suppl-KN-decomposition-2}
\end{align}
\end{subequations}

We now consider $\Kb_N^{(1)}$ and $\Kb_N^{(2)}$ as functions of $t_N$. By differentiating them over $t_N$ and using Eq.~\eqref{eq:Suppl-QBE}, we obtain the following equations of motion
\begin{subequations}
\label{eq:Suppl-KN-12-eom}
\begin{align}
\label{eq:Suppl-KN-12-eom-1}
\partial_{t_N} \Kb_N^{(1)} =&\, \Lambda_{\rm ens} \Big[\Kb_N^{(1)}  - \rb_{\rm st} K_{N-1} \Big],  \\
 \partial_{t_N} \Kb_N^{(2)} =&\,\Lambda_{\rm ens}\, \Kb_N^{(2)}. \label{eq:Suppl-KN-12-eom-2}
\end{align}
\end{subequations}
The initial condition for $\Kb_{N}^{(1)}$ is
\begin{align}
\label{eq:Suppl-KN-1-initial-cond}
& \Kb^{(1)}_N(t_{N-1})\equiv
\Kb^{(1)}_{\ell_1...\ell_N}(t_N=t_{N-1},t_{N-1},...,t_1) \nonumber \\
&\hspace{0.5cm}=\big\langle \rb_{N-1} \big(\nb_{\ell_{N-1}}\rb_{N-1}\big)\,I_{\ell_{N-2}}(t_{N-2})...I_{\ell_1}(t_1)\big\rangle,
\end{align}
and the initial condition for $\Kb_{N}^{(2)}$ can be obtained by averaging over the noise $\xi_{\ell_{N-1}}(t_{N-1})$ in the same way as in the main text (for the two-time correlator), that gives
\begin{align}
\label{eq:Suppl-KN-2-initial-cond}
& \Kb^{(2)}_N(t_{N-1})\equiv
\Kb^{(2)}_{\ell_1...\ell_N}(t_N=t_{N-1},t_{N-1},...,t_1)=
    \nonumber \\
&\big\langle \big[\nb_{\ell_{N-1}}- (\nb_{\ell_{N-1}} \rb_{N-1})\, \rb_{ N-1}+\mathcal{K}_{\ell_{N-1}}(\nb_{\ell_{N-1}} \times \rb_{N-1}) \big] \nonumber \\
&\hspace{0.2cm} \times I_{\ell_{N-2}}(t_{N-2})...I_{\ell_1}(t_1)\big\rangle.
\end{align}
We then solve the linear equations~\eqref{eq:Suppl-KN-12-eom} using  \eqref{eq:Suppl-rens-sol},
\begin{subequations}
\label{eq:KN-12-sol}
\begin{align}
\Kb^{(1)}_N =&\, \mathcal{P}(t_N|t_{N-1})\, \Kb^{(1)}_N(t_{N-1}) + \boldsymbol{\mathcal{P}}_{\rm st} K_{N-1},
    \\
\Kb^{(2)}_N =&\, \mathcal{P}(t_N|t_{N-1})\, \Kb^{(2)}_N(t_{N-1}),
\end{align}
\end{subequations}
and inserting the initial conditions \eqref{eq:Suppl-KN-1-initial-cond}--\eqref{eq:Suppl-KN-2-initial-cond}, we find
\begin{align}
& \Kb^{(1)}_N + \Kb^{(2)}_N  =\mathcal{P}(t_N|t_{N-1}) \times
    \nonumber\\
& \big\langle \big[\nb_{\ell_{N-1}} +\mathcal{K}_{\ell_{N-1}}(\nb_{\ell_{N-1}} \times \rb_{N-1}) \big] I_{\ell_{N-2}}(t_{N-2})...I_{\ell_1}(t_1)\big\rangle
    \nonumber \\
& + \boldsymbol{\mathcal{P}}_{\rm st}(t_N|t_{N-1})\, K_{N-1},
\label{eq:Suppl-K12}
\end{align}
where $K_N$ is the short notation for the correlator (\ref{eq:Suppl-KN}).

Finally, using Eqs.\ \eqref{eq:Suppl-KN-def}, \eqref{eq:Suppl-KN-vector}, and (\ref{eq:Suppl-KN-decomposition-0}), the result (\ref{eq:Suppl-K12}) can be rewritten as a recursion,
\begin{align}
\label{eq:Suppl-KN-recursive}
& \Kb_N = \mathcal{P}(t_N|t_{N-1})\big[\nb_{\ell_{N-1}} K_{N-2}
    \nonumber \\
& \hspace{0.5cm} +\mathcal{K}_{\ell_{N-1}}(\nb_{\ell_{N-1}} \times \Kb_{N-1}) \big]
 +\boldsymbol{\mathcal{P}}_{\rm st}(t_N|t_{N-1})\, K_{N-1},
\end{align}
which is exactly the same as Eq.\ (\ref{eq:Suppl-Kvec-N-recursive-1-GCR}) for the vector-valued correlators obtained via the GCR method [recall that Eq.\ (\ref{eq:Suppl-KN}) is also the same as Eq.\ (\ref{eq:Suppl-KN-alternative-form-GCR})].
It is easy to see that $\Kb_N$ in the initial cases $N=1$ and $N=2$ for the recursive relation (\ref{eq:Suppl-KN-recursive}) also coincide with the results \eqref{eq:Suppl-recursive-2-GCR} and (\ref{eq:Suppl-Kb2}) for the GCR method [so that we can still define $K_0=1$ as in Eq.\ (\ref{eq:Suppl-K0})]. This proves that $\Kb_N=\Kb_N^{\rm GCR}$, so any multi-time multi-detector correlator calculated via the generalized collapse recipe coincides with the correlator given by the quantum Bayesian formalism. The obvious advantage of the recipe is simplicity of calculations compared with the direct quantum Bayesian simulations.

Note that for a single detector ($N_{\rm d}=1$), the correlators can be larger than 1 only in the presence of a unitary evolution. This is because the projection of the collapsed state (\ref{eq:Suppl-coll-state}) on the measurement axis is $\pm 1$ (even though it is outside the Bloch sphere), and without unitary evolution (only decoherence) this projection remains within the $\pm 1$ range. In contrast, for detectors of non-commuting observables, the correlators can exceed 1 even without unitary evolution, only due to phase backaction. As an example, for continuous measurement of $\sigma_z$ and $\sigma_x$ \cite{Suppl-Shay2016}, the two-time cross-correlator $K_{zx}(t_1, t_2)$ exceeds 1 for small positive values of $t_1$ and $t_2-t_1$ if the initial state is $\rb(0)=(0,-1,0)$ and the phase backaction for $\sigma_z$-measurement is sufficiently strong, ${\mathcal K}_z=\tan \varphi_{z}^{\rm a}>1$. A weaker phase backaction would also produce cross-correlator larger than 1 if $\sigma_x$ measurement is replaced with the measurement along the direction between $x$ and $z$.

\end{document}